\documentclass[prb,aps,twocolumn,amsmath,amssymb,floatfix,superscriptaddress]{revtex4-1}
\usepackage{graphicx,amsmath}
\usepackage{float}
\usepackage{subfigure}
\usepackage{mathrsfs}
\usepackage{graphics}
\usepackage{epstopdf}
\usepackage{amssymb}
\usepackage{hyperref}
\usepackage{color}
\usepackage{braket}
\usepackage{multirow}
\begin{document}

\author{Kallol Mondal}
\email[E-mail: ]{kallolsankarmondal@gmail.com}
\affiliation{Physics and Applied Mathematics Unit, Indian Statistical Institute, 203 Barrackpore Trunk Road, Kolkata-700108, India}

\author{Sudin Ganguly}
\email[E-mail: ]{sudinganguly@gmail.com}
\affiliation{Department of Physics, School of Applied Sciences, University of Science and Technology Meghalaya, Ri-Bhoi-793101, India }

\author{Santanu K. Maiti}
\email[E-mail: ]{santanu.maiti@isical.ac.in}
\affiliation{Physics and Applied Mathematics Unit, Indian Statistical Institute, 203 Barrackpore Trunk Road, Kolkata-700108, India}

%\date{\today}
\title{Strain-induced thermoelectricity in pentacene}
\begin{abstract}
The present work discusses a non-synthetic strategy to achieve a favorable thermoelectric response in pentacene via strain. It is found that a uni-axial strain is capable of inducing spatial anisotropy in the molecule. As a result, the transmission spectrum becomes highly asymmetric under a particular strained scenario, which is the primary requirement to get a favorable thermoelectric response. Different thermoelectric quantities are computed for the strain-induced pentacene using Green's function formalism following the Landauer-B\"{u}ttiker prescription. Various scenarios are considered to make the present work more realistic, such as the effects of substrate, coupling strength between the molecule and electrodes, dangling bonds, etc. Such a scheme to enhance the thermoelectric performance in pentacene is technologically intriguing and completely new to the best of our knowledge.
\end{abstract}
%\pacs{05.45.Df, 72.15.Jf, 72.20.Pa, 05.60.Gg}
\maketitle
%%%%%%%%%%%%%%%%%%%%%%%%%%%%%%%%%%%%%%%%%%%%%%%%%%%%%%%%%%%%%%%
\section{\label{sec:intro}Introduction}

The detrimental nature of climate change led us to think of efficient, renewable, and unconventional energy sources. A recent study reveals that $63\%$ of the global energy consumption is wasted as a form of heat energy\cite{CULLEN20102059,doi:10.1126/science.1158899}. Naturally, it prompts us to develop efficient mechanisms which convert this `waste-heat' energy into electrical energy. Thermoelectric (TE) phenomenon enables direct conversion of heat energy into electricity and vice versa\cite{snyder2008,He-tritt-Science}. Therefore, thermoelectricity can potentially be one of the unconventional energy sources, which can also address the global energy crisis and global warming. At present, the wide applicability of TE devices hinges on the efficacy of the TE material.

 The efficiency of TE material is quantified by a dimensionless parameter, dubbed as \textit{figure of merit} ($ZT$), and expressed as~\cite{PhysRevB.47.12727}
\begin{equation}
ZT = \frac{G S^2 T}{k}
\label{eq:fom}
\end{equation}
where $S, \sigma, T$, and $k$ are the Seebeck coefficient, electrical conductance, temperature, and total thermal conductance,
respectively. Since both electrons and phonons are heat carriers, the total thermal conductance comprises of two parts -- electronic part ($k_\text{el}$) and phononic part ($k_\text{ph}$). Typically, a material with $ZT>1$ is regarded as a good TE material. However, to be economically competitive, materials with $ZT \sim 3$ are often recommended~\cite{Tritt-review}. To have larger TE efficiency of a material, a large Seebeck coefficient, high electrical conductivity, and low thermal conductivity are highly solicited. However, it turns out that these material properties are interrelated and also possess conflicting interests. For example, to get higher $ZT$, we want large electrical transport and, at the same time, suppression of the phonon transport which is conflicting~\cite{Zhang-Adv-Mat}. These contrasting attributes pose a great challenge to the optimization of the functional element as a TE material.

In recent years, there has been a significant advancement in the TE sector~\cite{Dresselhaus2007}, and it has been identified that nanostructures and low-dimensional systems show better performance than the bulk ones~\cite{PhysRevLett.100.066801,PhysRevB.47.12727,PhysRevB.47.16631}. But, the technology to obtain such structures is complicated, expensive, and hence far from commercialization. Recent advancements in this field indicate that there is a growing inclination towards the organic compound as a TE material. The organic materials enjoy superiority over the inorganic ones due to their unique characteristics of being flexible, low-cost, diversiﬁed molecular design, ease of manufacturing, light-weight, and above all, low intrinsic thermal conductivity, typically below $1 \rm{W m}^{ -1} \rm{K}^{ -1}$ ~\cite{Zhang2020,Russ2016,Zhang-Adv-Mat}. 

Of late, the idea of using single molecules as a functional element has propelled extensive experimental and theoretical investigations to understand the charge transport through molecular junctions~\cite{cuevas2010molecular}. The molecular junctions are thought to be promising candidates to deliver high efficiencies due to the discreteness of the energy spectrum, and the properties can be tuned via chemical synthesis, electrostatic gates, or by applying pressure~\cite{C6CS00141F}. Motivated by these, pentacene which is an organic molecule, consists of five linearly-fused benzene rings, is chosen as the TE material for the present work. However, for a pure pentacene molecule, $ZT$ is close to one~\cite{wang2009first}, which is still far beyond the level of meeting the practical applications. Nevertheless, significant progress has been made in developing high-performance TE devices using organic molecules by employing various strategies like chemical doping~\cite{KANG2019112,Kang2016,Yamashita2019}, field-modulated doping~\cite{Huang-Ange,Pernstich2008,Zhang-adv-mat-2015}, and strain~\cite{ZZhang2021}.

It turns out that carrier mobility is one of the key factors to determine the TE response. For small organic molecules like pentacene, the inter-molecular packing and molecular structure become decisive to the carrier mobility~\cite{da-silva-admat, yuan2014ultra} and hence the TE performance. Due to the presence of weak Van der Waals interaction, these organic molecules easily undergo lattice deformation under the application of external force~\cite{yuan2014ultra,Reyes-Martinez2015}. Thus, as a non-synthetic strategy, introducing strain in such molecular junction could be an efficient way to improve the TE performance~\cite{ZZhang2021}. For example, in case of TIPS (tri-isopropylsilyly-ethynyl) pentacene, carrier mobility increases from $0.8$ to $4.6\,$cm$^2$V$^{-1}$s$^{-1}$ when the $\pi - \pi$ stacking distance is decreased from $3.33\,$\AA ~to $3.08\,$\AA~\cite{Giri2011}. There are other studies where the directional dependence of TE properties on strain is explored~\cite{C7MH00489C,Landi-adv-mat}. However, to the best of our knowledge, the effect of strain on the TE properties of the organic molecules is yet to be fully understood.

In this present communication, we study the TE response in a typical small organic molecule like pentacene and propose a non-synthetic strategy to achieve favorable TE performance. The strain is introduced along the direction of the benzene chain, and its effect is incorporated through the hopping integral~\cite{Ribeiro_2009,Naumis_2017}. The strain effectively changes different bond lengths depending upon the strain direction and the corresponding hopping integrals get modified accordingly. This in turn induces a spatial anisotropy in the hopping terms of the molecule. Consequently, an asymmetry in the transmission spectrum is expected, which is the key requirement to get higher $ZT$~\cite{Mahan7436,PhysRevApplied.3.064017}. We employ the non-equilibrium Green's function formalism, based on Landauer-B\"{u}ttiker prescription to compute two-terminal transmission probability~\cite{datta1997electronic, datta2005quantum}. Using Landauer's prescription, the TE performance is examined by evaluating the electrical conductance, thermopower, and thermal conductance due to both the electrons and phonons~\cite{PhysRevApplied.3.064017,PhysRevB.79.033405}. In the case of small organic molecules, at low temperature, the contribution of phonon to the thermal conductance is very small compared to the electronic one and hence can be safely ignored~\cite{PhysRevB.102.245412}. However, at high temperatures, such as room temperature, the phonon thermal conductance must be included.

The key findings of our work are (i) single organic molecules can be a promising class of TE materials, (ii) our study provides a nonsynthetic strategy for optimizing TE properties of organic molecules, (iii) \textit{figure of merit} of pure pentacene molecule can be significantly enhanced by applying strain, and (iv) compared to stretching, compression along the benzene chain shows a favorable TE response.

The rest of the presentation is organized as follows. In Sec. II, we present the model and the necessary theoretical formulation to compute different TE quantities. All the results are critically analyzed in Sec. III. Finally, in Sec. IV, we end with concluding remarks.

%%%%%%%%%%%%%%%%%%%%%%%%%%%%%%%%%%%%%%%%%%%%%%%%%%%%%%%%%%%%%%%

\section{\label{sec:formalism}Theoretical formulation}

\begin{figure}[t!]
\centering
\boxed{\includegraphics[width=0.45\textwidth]{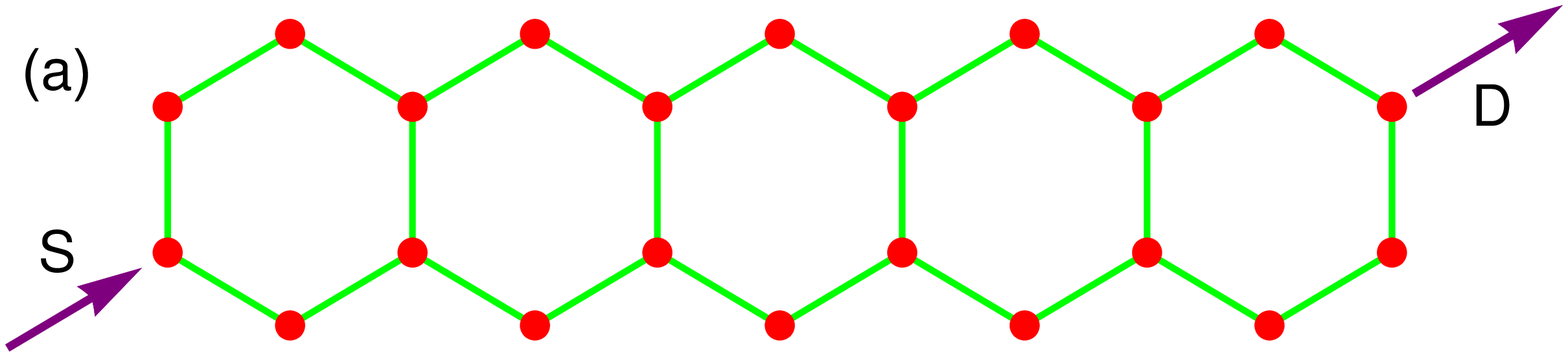}} \\
\boxed{\includegraphics[width=0.45\textwidth]{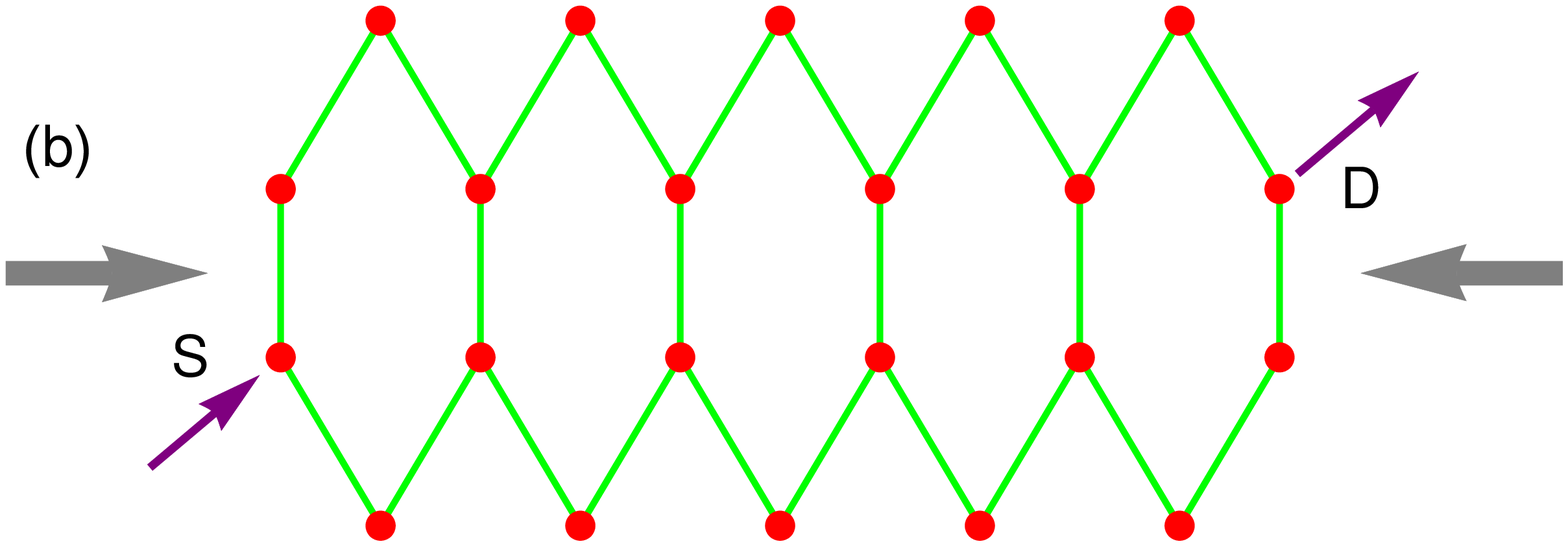}} \\
\boxed{\includegraphics[width=0.45\textwidth]{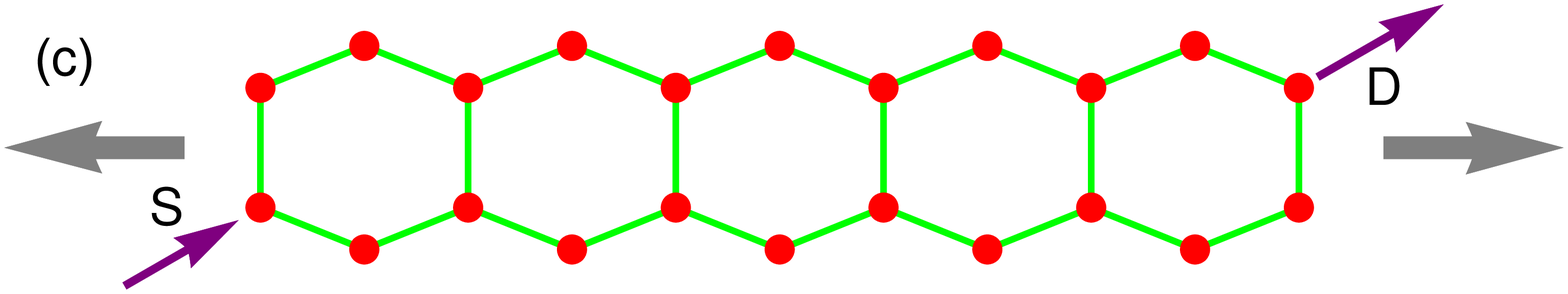}} 
\caption{(Color online). Schematic view of the pentacene molecular junction which is  connected in between two 1D semi-infinite electrodes, namely source (S) and drain (D), kept at two different temperatures, $T + \Delta T/2$ and $T- \Delta T/2$, where $\Delta T$ is infinitesimally small. The red circles denote the atomic sites in the pentacene. The purple arrows denote the source and drain respectively. (a) In the absence of strain. In presence of a uni-axial strain, applied along the chain of the benzene rings with (b) compressed and (c) stretched scenarios. The grey arrows indicate the stretching and the squeezing of the molecular junction, accordingly.}
\label{fig:diagram}
\end{figure}

Figure~\ref{fig:diagram} depicts the schematic diagram of our proposed device where a single pentacene molecule is connected with two one-dimensional (1D) semi-infinite electrodes, namely source ($S$) and drain  ($D$), denoted by purple arrows. These two electrodes are kept at two different temperatures $T + \Delta T/2$ and $T- \Delta T/2$, where $\Delta T$ is infinitesimally small. Thus, we restrict ourselves to the linear response regime. Figure~\ref{fig:diagram}(a) shows the strain less case while Figs.~\ref{fig:diagram}(b) and (c) show the schematic diagrams in presence of strain. We use the nearest-neighbor tight-binding Hamiltonian to describe the system.   

%################################################################
%%%%%%%%%%%%%%%%%%%%%%%%%%%%%%%%%%%%%%%%%%%%%%%%%%%%%%%%%%%%%%%

\vspace{0.5cm}
\noindent \textbf{Model Hamiltonian:} 
The  Hamiltonian $H$ representing the total system comprises of four parts and it is written as
\begin{equation}
H = H_\text{P}  + H_\text{S}  + H_\text{D}  +H_\text{C}.
\end{equation}
Where $H_\text{P}, H_\text{S}, H_\text{D},$  and $ H_\text{C} $ are the sub-parts of the Hamiltonian associated with the molecular junction, source, drain, and the coupling between the electrodes and the molecular junction, respectively. Each of these sub-parts of the Hamiltonian is given by
\begin{subequations}
\begin{eqnarray}
H_\text{P} & = & \sum_n \epsilon_n c_n^\dagger c_n + \sum_{\langle nm \rangle}  \left(t_{nm} c_n^\dagger c_m + h.c. \right), \\
H_{\rm S} &=& H_{\rm D} = \epsilon_0\sum\limits_{n} d_n^{\dagger} d_n +
t_0\sum\limits_{\langle nm\rangle}\left(d_n^{\dagger} d_m + h.c.\right),\\
H_{\rm C}  &=& H_{\rm S,\rm P} + H_{\rm D, \rm P} \nonumber \\
& = & \tau_S\left(c_p^{\dagger} d_0 + h.c.\right) + \tau_D\left(c_q^{\dagger} d_{N+1} + h.c.\right).
\end{eqnarray} 
\end{subequations}
where $\epsilon_n$ denotes the on-site energy at site $n$ and $t_{mn}$ represents hopping amplitude from site $m$ to $n$. Here, $c,d$ represent the annihilation operators whereas their hermitian conjugates $c^\dagger, d^\dagger$ denote the creation operators and satisfy  anti-commutation relation. The parenthesis $\langle .. \rangle$ denotes that the interaction is limited to nearest-neighbor only.  For pure pentacene molecule, on-site energies are independent of $n$ and can be taken as a constant.  On-site potential $\epsilon_0$ and hopping amplitude $t_0$  are taken to be the same for both the electrodes. The coupling between the source(drain) and the molecular junction is denoted by $\tau_S(\tau_{D})$. 

\vspace{0.5cm}
\noindent \textbf{Incorporation of strain:} Here, we consider a uni-axial strain, applied along the chain of the benzene rings. The introduction of strain changes the distances between atomic sites, and thus the hopping amplitude is modified accordingly. The modified hopping integral with the bond length is given by~\cite{Ribeiro_2009,Naumis_2017}

\begin{equation}
t_{nm} = t_0 ~e^{- \beta\left(\frac{l}{a_0}-1\right)}
\label{strain}
\end{equation}
where $t_0$ is the hopping integral without strain, while $t_{nm}$ is the modified hopping inegral due to strain. $a_0$ and $l$ be the lattice constant and modified bond length respectively. $\beta$ is the Gr\"{u}neisen parameter, a measure of the phonon mode softening or hardening rate. This parameter can be calculated using Raman scattering experiment~\cite{PhysRevB.79.205433,Ding2010} or from \textit{ab initio} calculations~\cite{PhysRevB.83.115449}. Usually, for graphene, Gr\"{u}neisen parameter ranges between 2-3. In a recent DFT-based work~\cite{Abdulla_2015}, Gr\"{u}neisen parameter for tetracene has been reported as 2.82. In such degree, we assume the Gr\"{u}neisen parameter $\beta\sim 3$ for pentacene in the present work.

We define a dimensionless quantity $\delta=\beta\left(\frac{l}{a_0}-1\right)$, which can be thought of as a continuous parameter to illustrate the effect of strain. Clearly, when $l > a_0$, the parameter $\delta$ becomes positive implying the stretching of the molecular junction. On the other hand, $\delta < 0$ indicates the compression of the same. Since we are introducing the strain along the benzene chain, only the angular hopping amplitude is modified, whereas the vertical hopping amplitude remains the same. 

It is to be noted that the functional form of the hopping integral under strain is taken from Ref.~\cite{Ribeiro_2009, Naumis_2017} which deals with the similar kinds of 2D structures like graphene where the geometrical building block is pentacene. Here we must mention that we aim to examine the effect of strain on the TE properties of pentacene which is invoked through the modification of hopping integral. The presence of strain modifies the distances between atomic sites, and the hopping amplitude is altered accordingly. It turns out that the modification of the hopping integral does influence the TE properties, as the transmission spectra become highly asymmetric, which is the primary requirement for favorable TE response. The specific functional form of $t_{nm}$, as mentioned in Eq.~\ref{strain}, is not so much important. One can also consider other functional forms as well, which we confirm through our detailed numerical calculations. The central issue is that we need to consider a finite (slow) variation of $t_{nm}$ in presence of strain. 

 In order to study the TE properties under strained and unstrained conditions, we need to calculate different TE quantities along with transmission function. The prescription is as follows.
%-----------------------------------------------------------------------------------------------------------------------------------------------------------------------------

\vspace{0.5cm}
\noindent \textbf{Two-terminal transmission probability:} We calculate the two-terminal transmission probability for electrons travelling from source to drain through the molecular junction. We employ the non-equilibrium Green's function technique~\cite{datta1997electronic,datta2005quantum} to evaluate the two-terminal transmission probability.  The retarded Green's function can be obtained from the following equation
\begin{equation}
\mathcal{G}^r = \left(E- H_{\text{P}}- \Sigma_S -\Sigma_D\right)^{-1}
\end{equation}
where $\Sigma_S$ and $ \Sigma_D$ represent the self-energies of the source and drain, respectively. Now, the transmission probability can be expressed in terms of retarded ($\mathcal{G}^r$) and advanced $\left(\mathcal{G}^a\left(=\mathcal{G}^r\right)^\dagger\right)$ Green's functions as
\begin{equation}
\mathcal{T}= \text{Tr}\left[\Gamma_S \mathcal{G}^r \Gamma_D \mathcal{G}^a \right]
\label{eq:trans}
\end{equation}
where $\Gamma_S$ and $\Gamma_D$ are the coupling matrices that describe the rate at which particles scatter between the electrodes and the molecular junction.

\vspace{0.5cm}
\noindent \textbf{Thermoelectric quantities:}  In the linear response regime, all the TE quantities like $G$, $S$, and $K_\text{el}$ can be extracted using Landauer’s integrals as~\cite{PhysRevApplied.3.064017,PhysRevB.79.033405,Mondal2021}

\begin{subequations}
 \begin{eqnarray}
 G & = & \frac{2 e^2}{h} L_0 \\
 S & = & -\frac{1}{e T }\frac{L_1}{ L_0}\\
k_\text{el} & = & \frac{2}{hT}\left( L_2 - \frac{L_1^2}{L_0}\right)
 \end{eqnarray}
  \label{eq:TE-quantity}
\end{subequations}
where the Landauer’s integral is given by
\begin{equation}
L_n = - \int   \mathcal{T}(E) (E- E_F)^n\left(\frac{\partial f}{\partial E}\right) dE.
\end{equation}

Where $h,f$, and $E_F$ indicate Planck's constant, equilibrium Fermi-Dirac occupation probability, and Fermi energy respectively. Here, $\mathcal{T}(E)$ is the two-terminal transmission probability. 

The \textit{figure of merit} can be expressed in the following form~\cite{Ganguly_2020,PhysRevApplied.3.064017}
\begin{equation}
ZT = \frac{G S^2 T}{k (=k_\text{el}+ k_\text{ph})}.
\end{equation}

The product of the electrical conductance and the square of the thermopower is known as the power factor i.e. $PF = G S^2 $. For a good TE material, $PF$ should be as high as possible. We must mention that the maximum power of a TE device not only depends on the power factor but also on the temperature difference between the electrodes. The expression for the maximum power is given by~\cite{PhysRevLett.106.230602,Briones-Torres2021}
\begin{equation}
P_{max} = \frac{1}{4} G S^2 (\Delta T)^2.
\label{eqn:maxP}
\end{equation} 

The efficiency of the TE device at the maximum power is given by~\cite{PhysRevLett.106.230602,Briones-Torres2021} 
\begin{equation}
\eta\left(P_{max}\right) = \frac{\eta_C}{2} \frac{ZT}{ZT+2} 
\label{eqn:etamaxP}
\end{equation}
where $\eta_{C}$ indicates the ideal Carnot efficiency.

\vskip 0.1 cm
\noindent{\bf Phonon thermal conductance:} 

Since we are working with a single-molecule, the lattice vibration is quite small at low temperature and hence $k_{\rm ph}$ is very small compared to its electronic counterpart~\cite{PhysRevB.102.245412}. However, that is not the case at high temperatures and hence the phononic contribution to the thermal conductance should be included to get a precise estimation of $ZT$. The phonon thermal conductance can be computed from the following expression~\cite{doi:10.1080/10407790601144755,PhysRevB.80.201408,aghosh} as
\begin{equation}
k_{ph}= \frac{\hslash}{2\pi}\int_0^{\omega_c} \mathcal{T}_{ph}\frac{\partial f_{BE}}{\partial T}\omega d\omega .
\end{equation}
Here, $\omega$ and $\omega_c$ are the phonon and cut-off frequencies respectively. We consider only elastic scattering in the present case. $f_{BE}$ denotes the Bose-Einstein distribution function. $\mathcal{T}_{ph}$ is the phonon transmission probability across the pentacene, evaluated using the standard Green's function technique as~\cite{aghosh}
\begin{equation}
\mathcal{T}_{ph}= \text{Tr}\left[\Gamma_S^{ph} \mathcal{G}_{ph} \Gamma_D^{ph} \left(\mathcal{G}_{ph}\right)^\dagger \right]
\end{equation}
$\Gamma_{S/D}^{ph}=i\left[\widetilde{\Sigma}_{S/D}-\widetilde{\Sigma}_{S/D}^\dagger\right]$ is known as the thermal broadening. $\widetilde{\Sigma}_{S/D}$ is the self-energy matrix for the source/drain electrode. The Green's function for pentacene reads as~\cite{aghosh}
\begin{equation}
G_{ph} = \left[{\mathbb M}\omega^2 - {\mathbb K} -\widetilde{\Sigma}_S - \widetilde{\Sigma}_D\right]
\end{equation}
where ${\mathbb M}$ is a diagonal matrix describes the mass matrix of pentacene. Each element of the mass matrix ${\mathbb M}_{nn}$ denotes the mass of the atom at the $n$-th position in the pentacene molecule. ${\mathbb K}$ is the matrix of spring constants in the pentacene. An element in the diagonal ${\mathbb K}_{nn}$ denotes the restoring force of the $n$-th atom due to its neighboring atoms, while the element ${\mathbb K}_{nm}$ represents the effective spring constant between $n$-th and $m$-th neighboring atoms. The self-energy matrices $\widetilde{\Sigma}_S$ and $\widetilde{\Sigma}_D$ have the same dimension as ${\mathbb M}$ and ${\mathbb K}$ and can be computed by evaluating the self-energy term $\Sigma_{S/D}=-K_{S/D}\,\text{exp}\left[2i\,\text{sin}^{-1}\left(\frac{\omega}{\omega_c}\right)\right]$, where $K_{S/D}$ is the spring constant at the electrode-pentacene contact interface.

Different spring constants are assumed through the second derivative of Harrison's interatomic potential~\cite{harrison}. For the 1D electrodes, we need to know the interatomic spacing $d$ and the elastic constant $c_{11}$, as transverse interaction is absent in 1D system. The expression for the spring constant is $K=3dc_{11}/16$. However, as the pentacene is a 2D object, the transverse interaction needs to be included through the other elastic constant $c_{12}$ and the expression for the spring for the pentacene becomes~\cite{Kittel2004} $K=3d\left(c_{11}+2c_{12}\right)/16$. Once the mass and spin constant are known, the cut-off frequency for the 1D electrodes is determined from the relation $\omega_c=2\sqrt{K/M}$. To have a detailed description of the methodology and to compute phonon thermal conductance, see Refs.~\cite{doi:10.1080/10407790601144755,PhysRevB.80.201408,aghosh}.

%%%%%%%%%%%%%%%%%%%%%%%%%%%%%%%%%%%%%%%%%%%%%%%%%%%%%%%%%%%%%%%%%%%%%%

\section{\label{sec:result}Numerical results and discussion}

In this section, we report the details of the TE response in pentacene with and without strain. We commonly refer $\delta=0$ case as the pristine molecule. To begin with, let us first mention the parameters which are kept fixed throughout the present work.  The lattice constant $a_0$ of pentacene is taken to be unity. The on-site energies of the molecular junction as well as the electrodes are set at zero. It is easy to establish that any non-zero values of on-site energy will not alter the essential physics of the present work. All the results reported here involving energies are measured in units of eV. We restrict ourselves in the wideband limit, and the NNH integral $t_0$ for the electrodes is kept fixed at $2\,$eV whereas for pentacene it is $1\,$eV. The coupling between the electrodes and the functional element ($\tau_S$ and $\tau_D$) are taken to be the same and are set at $0.75\,$eV. 

%################################################################
\begin{figure}[h!]
\centering
\includegraphics[width=0.238\textwidth]{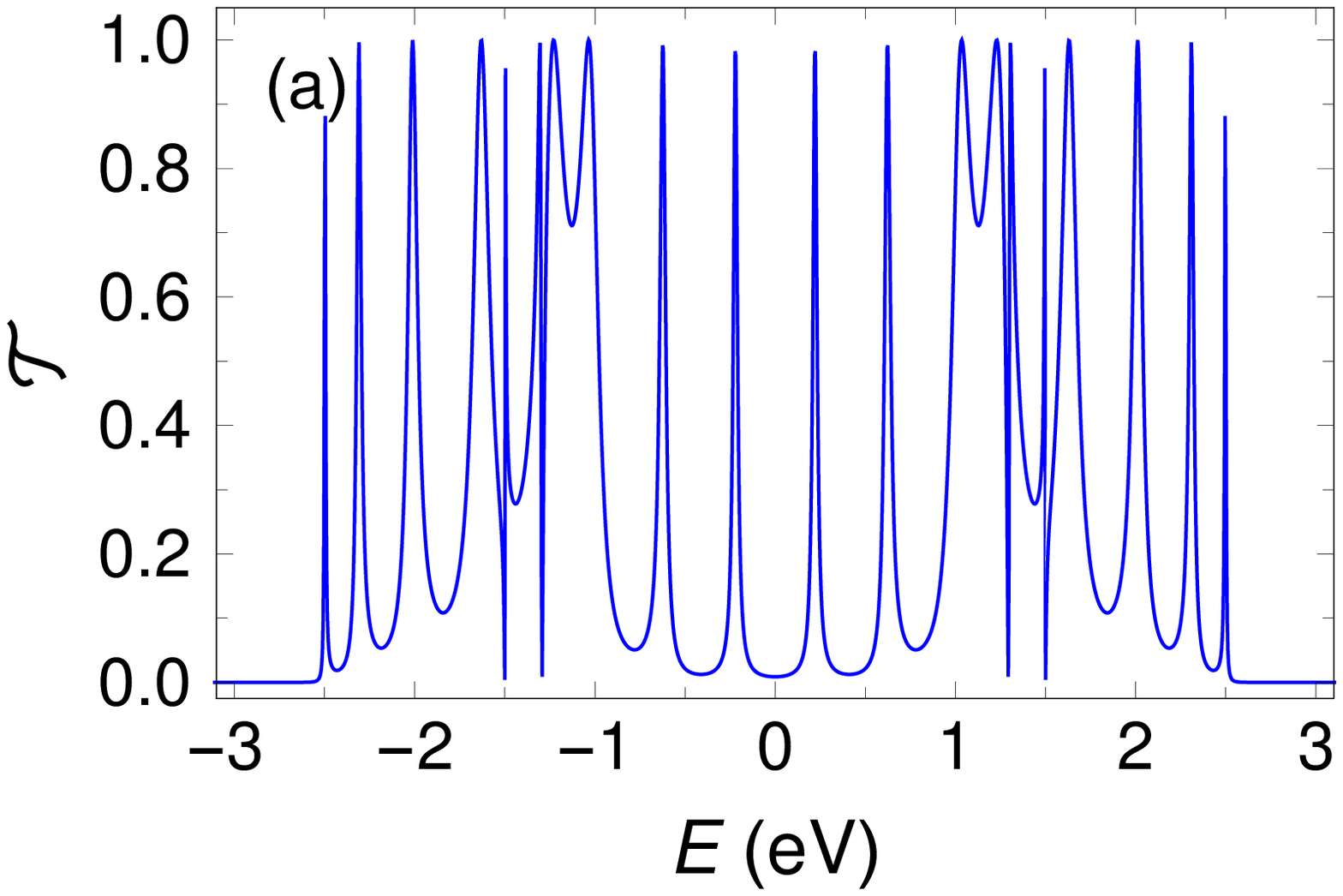} \hfill
\includegraphics[width=0.238\textwidth]{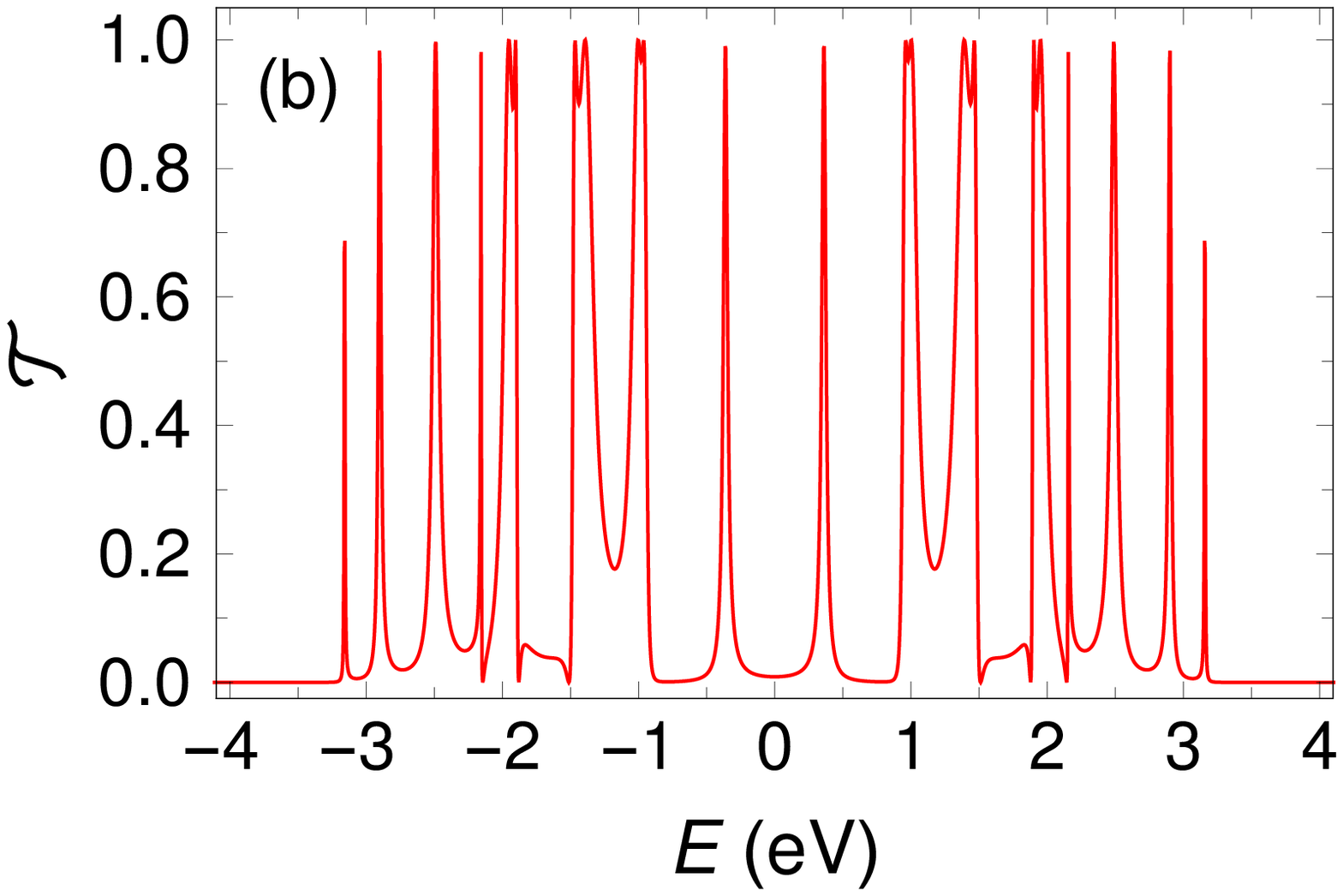} \vskip 0.1in
\includegraphics[width=0.238\textwidth]{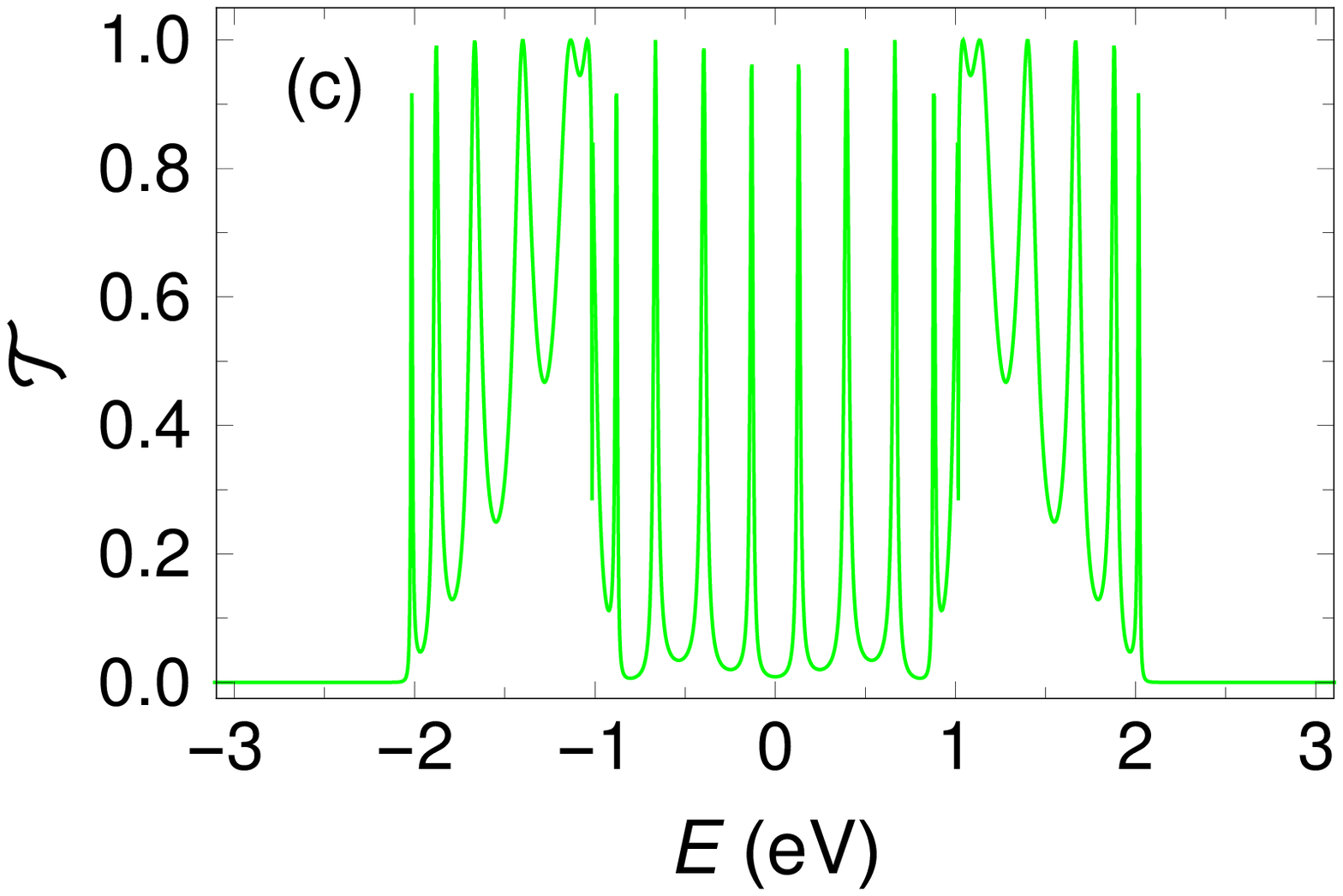} \hfill
\includegraphics[width=0.238\textwidth]{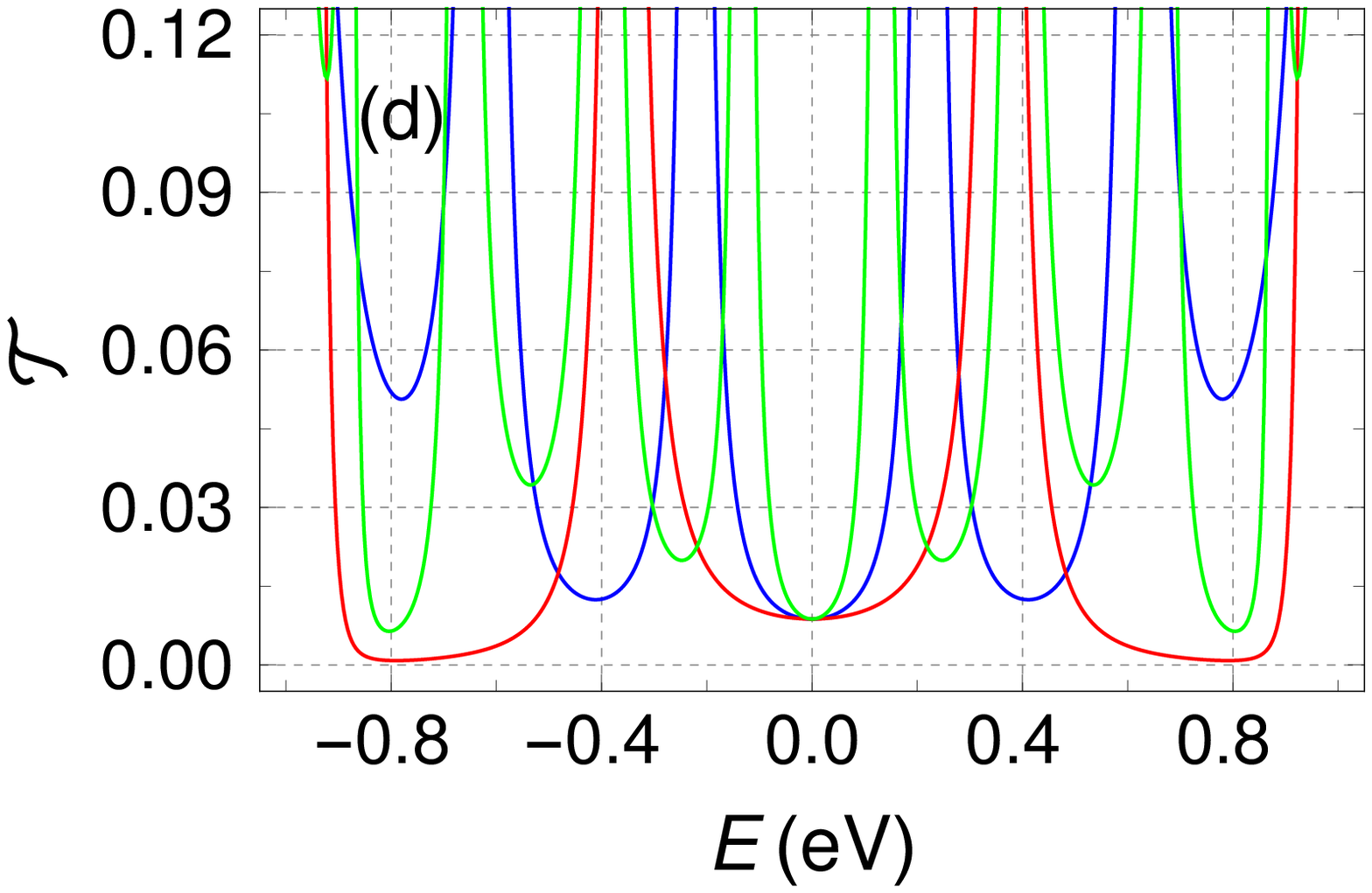} 
\caption{(Color online). Two terminal transmission probability ${\mathcal T}$ as a function of energy $E$. (a) In the absence of strain ($\delta =0$) denoted with blue color. In presence of uni-axial strain with (b) $\delta = 0.3$ (red color) and (c) $\delta = -0.3$ (gree color). (d) Transmission spectra for the former three cases within a narrow energy window to get a clear view around the band center (viz. $E=0$).}
\label{fig:trans}
\end{figure}
%%%%%%%%%%%%%%%%%%%%%%%%%%%%%%%%%%%%%%%%%%%%%%%%%%%%%%%%%%%%%%%
%################################################################
\begin{figure*}[t!]
\centering
\includegraphics[width=0.32\textwidth,height=0.2\textwidth]{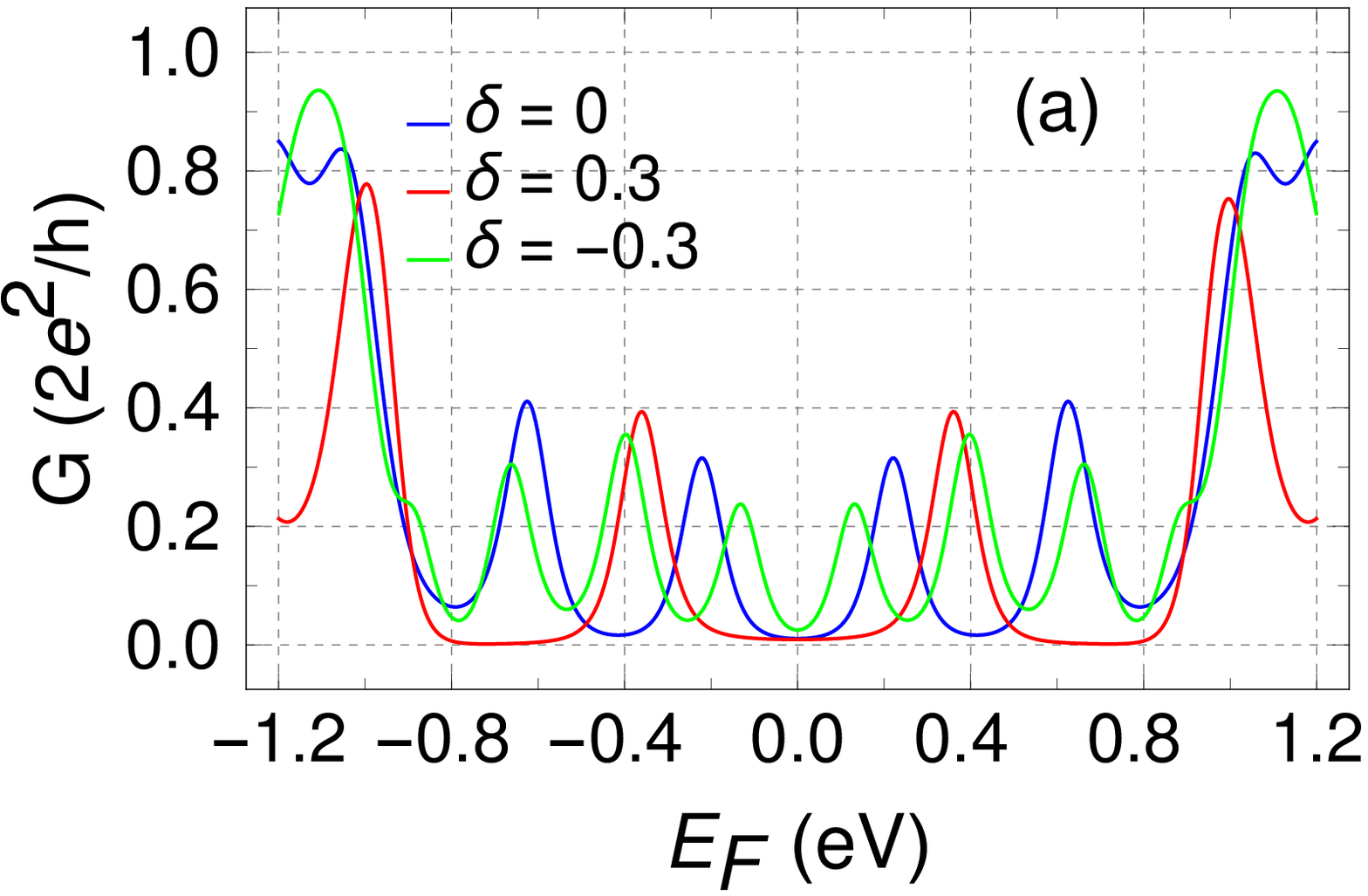}
\includegraphics[width=0.32\textwidth,height=0.2\textwidth]{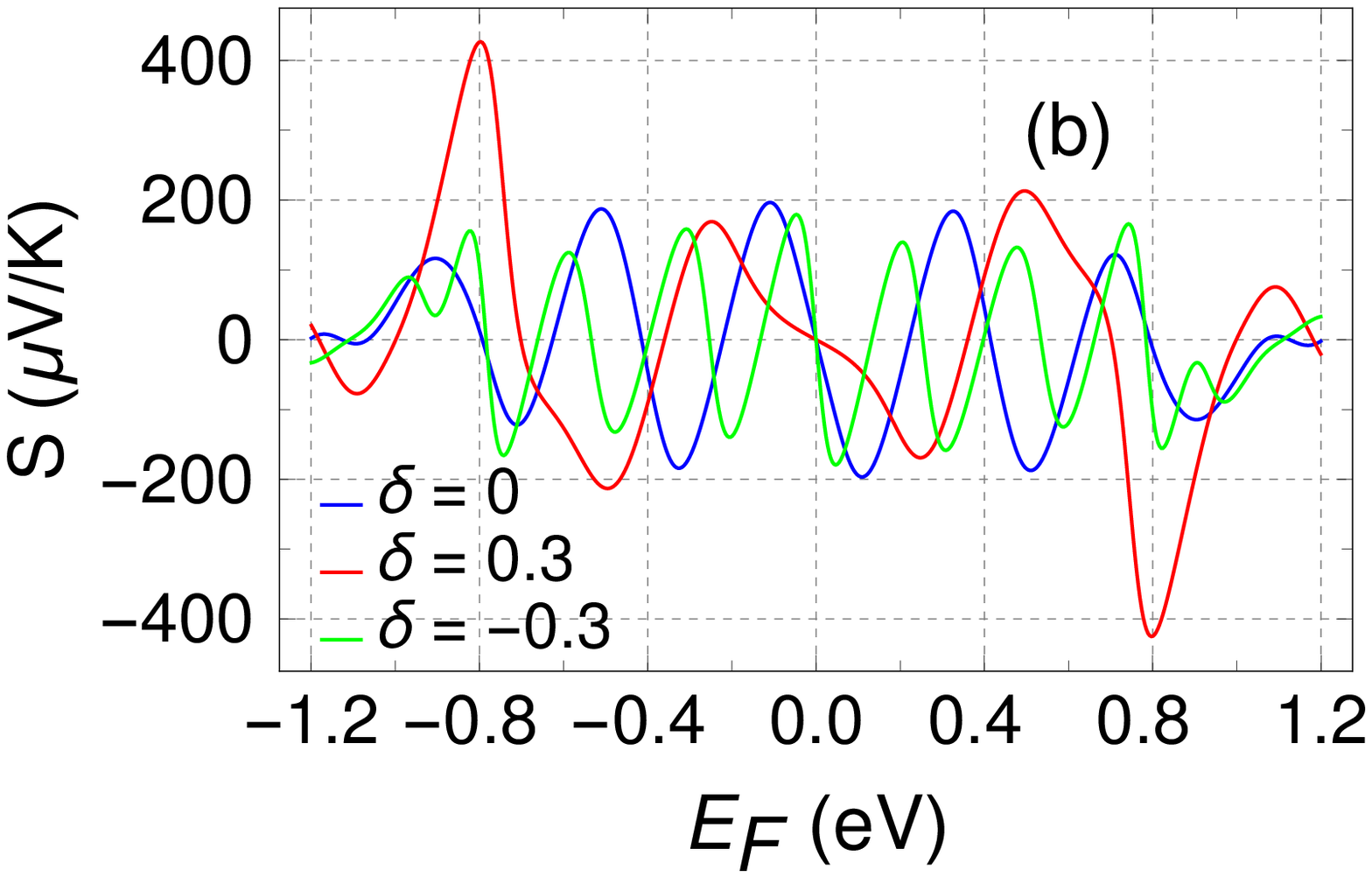} 
\includegraphics[width=0.32\textwidth,height=0.2\textwidth]{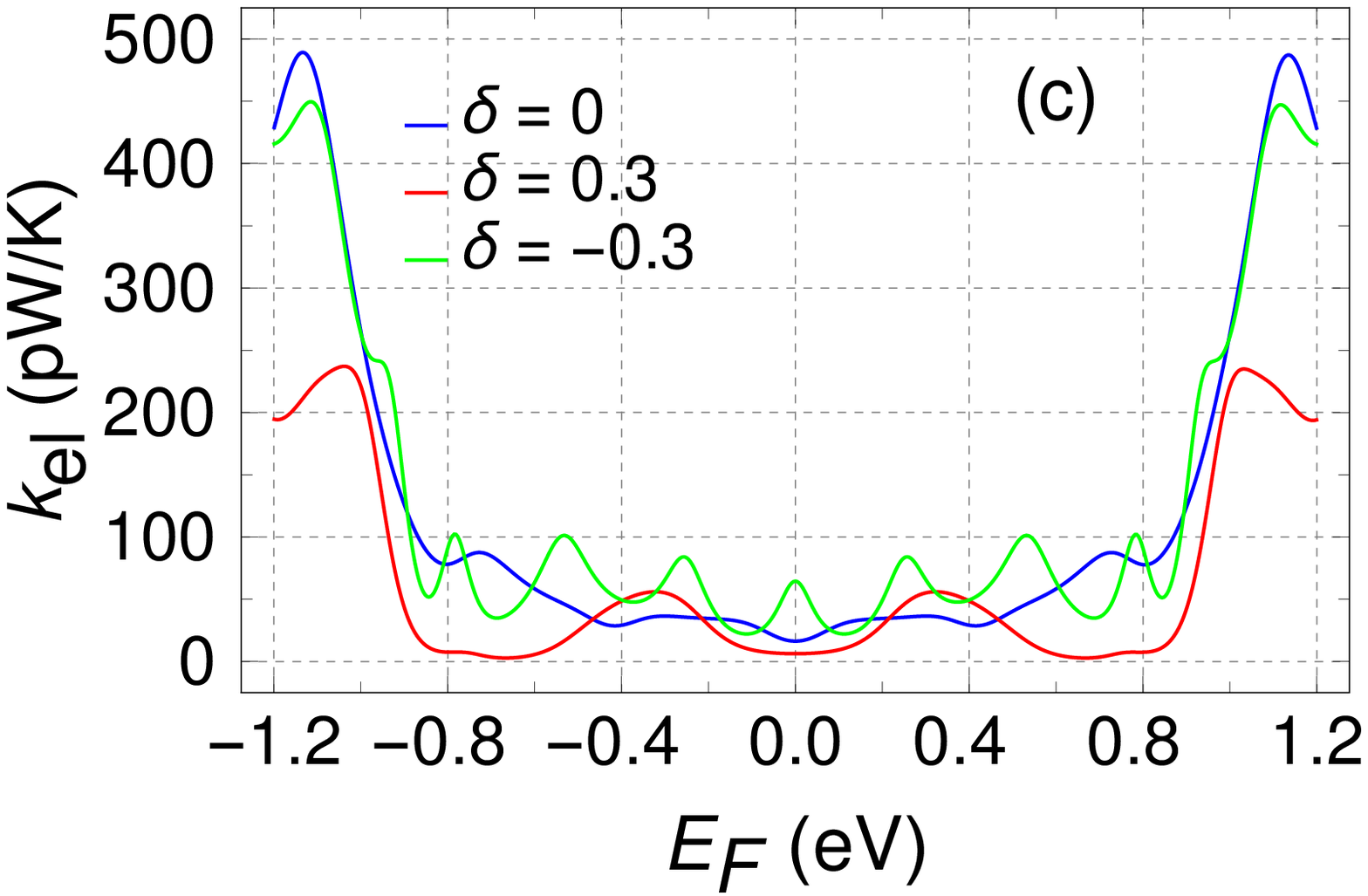} 
\caption{(Color online). Behavior of different thermoelectric quantities with Fermi energy $E_F$ at room temperature. (a) Electrical conductance $G$, (b) thermopower $S$, and (c) thermal conductance $k_{\rm el}$. The results corresponding to $\delta=0$, 0.3, and -0.3 are denoted with blue, red, and green colors, respectively.}
\label{fig:TEall}
\end{figure*}
%%%%%%%%%%%%%%%%%%%%%%%%%%%%%%%%%%%%%%%%%%%%%%%%%%%%%%%%%%%%%%%
Before presenting the results, it is imperative to have an estimation of the modified bond lengths under strained conditions. We compute the results for typical cases of the strained molecule, namely, $\delta=0.3$ and $\delta=-0.3$, for illustrations. When $\delta=0.3$, the angular bond length is stretched by 10$\%$ to $a_0$, with assumed Gr\"{u}neisen parameter $\beta=3$. Whereas, the angular bond length is compressed by the same amount as the former case when $\delta=-0.3$.  
\vskip 0.2 cm
\noindent{\bf Transmission Spectra:}

We begin our discussion by analyzing the two-terminal transmission probability ($\mathcal{T}$) as a function of energy both in the absence and presence of strain. The transmission probability is calculated following Eq.~\ref{eq:trans}. The variation of $\mathcal{T}$ as a function of energy for three cases, namely, $\delta =0$ and $\delta = \pm 0.3$ are shown in Fig.~\ref{fig:trans}. In Fig.~\ref{fig:trans}(b), 
for $\delta=0.3$, the allowed energy window is wider than that of the pristine case (Fig.~\ref{fig:trans}(a)), while it gets shortened for $\delta=-0.3$ (fig.~\ref{fig:trans}(c)). Such modifications in the allowed energy window are completely due to the modified hopping integrals under different strained conditions as is expected from Eq.~\ref{strain}. The transmission spectra are also significantly modified in the presence of strain. Interestingly, in Fig.~\ref{fig:trans}(b), we see that on either side of $E=0$, the transmission probability is vanishingly small over a narrow energy window associated with a sharp peak. This is clearly visible in Fig.~\ref{fig:trans}(d), where the transmission spectra are shown within a narrow energy window for the former three $\delta$-values. Such a situation indicates that the transmission spectrum is asymmetric over that energy region and may yield a favorable TE response~\cite{Ganguly_2020} as we shall see in the forthcoming discussion. It should be noted here that the asymmetry in the transmission function occurs due to the induced spatial anisotropy in the hopping integrals by strain.

\textit{Thermoelectric quantities:} Now, let us study the thermoelectric response of pentacene by analyzing the TE entities like electrical conductance, thermopower,  thermal conductance, and \textit{figure of merit} at room temperature ($T=300\,$K). 
Two different strained scenarios are considered, namely, $\delta=0.3$ and -0.3, including the pristine case, that is $\delta=0$ and the corresponding results are represented by blue, red, and green colors, respectively as shown in Fig.~\ref{fig:TEall}. 

The variation of electrical conductance $G$ (in units of $2e^2/h$) with Fermi energy $E_F$ is shown in Fig.~\ref{fig:TEall}(a). Within the given Fermi energy window, the behavior of $G$ is distinctly different for the three considered $\delta$-values, where the sharp peaks in their respective transmission spectrum (Fig.~\ref{fig:trans}) are smeared out due to the temperature broadening. At and around $E_F\approx \pm 0.6$, $G$ is vanishingly small for the stretched condition ($\delta=0.3$, red curve) while it is finite for the other two cases ($\delta=0$ and -0.3) within the said Fermi energy region. 

The behavior of thermopower $S$ as a function of Fermi energy is depicted in Fig.~\ref{fig:TEall}(b). Interestingly, $S$ is about 400$\,\mu$V/K at $E_F\sim \pm 0.8$ for $\delta =0.3$, while for the other two $\delta$-values, $S$ is restricted within $\pm 200\,\mu$V/K within the given Fermi energy window. Such a significantly enhanced $S$ for the stretched condition is completely due to the asymmetry in the corresponding transmission spectrum as noticed earlier in Fig.~\ref{fig:trans}. This in turn may increase the {\it figure of merit}, since $ZT\propto S^2$. 

Like the electronic conductance, the thermal conductance due to electrons $k_{\rm el}$ shows more or less similar behavior as a function of Fermi energy as shown in Fig.~\ref{fig:TEall}(c). We see that $k_{\rm el}$ attains relatively lower values for the stretched condition ($\delta=0.3$) than the other two cases throughout the given Fermi energy window. Moreover, $k_{\rm el}$ is significantly suppressed for $\delta=0.3$  close to the band center and also at and around $E_F\sim 0.8$. Since $k_{\rm el}$ appears in the denominator in the expression of $ZT$, a stretched scenario should yield a favorable TE response.    

What we accumulate from Figs.~\ref{fig:TEall}(a-c) is that the introduction of strain significantly modifies the different TE quantities. Particularly, for $\delta=0.3$ case, though the electrical conductance is low but larger thermopower and significantly suppressed thermal conductance due to electrons should result in a better TE response. The combined effect of all these TE quantities will be reflected in the behavior of $ZT$ which we will discuss now.

\begin{figure}[ht!]
\centering
\includegraphics[width=0.495\textwidth]{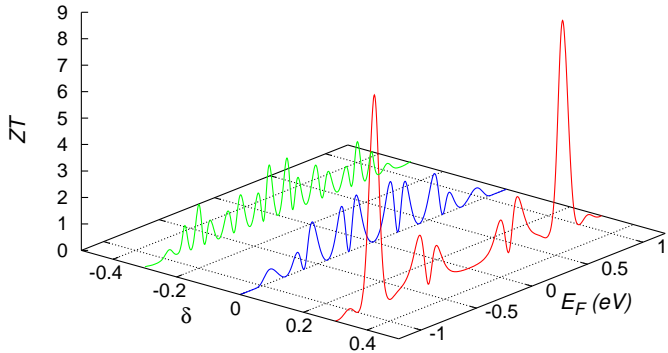} 
\caption{(Color online). $ZT$ at room temperature as a function of Fermi energy $E_F$ for three different strain parameters $\delta = 0$ and $\pm 0.3$. The color conventions are identical as described in Fig.~\ref{fig:TEall}.}
\label{fig:fom}
\end{figure}

The surface plot in Fig.~\ref{fig:fom} shows the variation of $ZT$ as a function of Fermi energy corresponding to the $\delta$-values as discussed earlier. For the pristine case, the maximum $ZT$ is found to be $\sim 2$ at room temperature which matches quite well with the existing \textit{ab initio} results, where $ZT$ of pure pentacene molecule is reported to be $0.8 - 1.8$ at $294\,$K~\cite{wang2009first}.  The introduction of a uni-axial strain significantly modifies the TE performance. For $\delta=-0.3$, TE response is not up to the mark as is seen by the green curve in Fig.~\ref{fig:fom}. The most favorable response is obtained for the stretched scenario with the strain parameter $\delta =0.3$, as is noticed by the two red 
%################################################################
%%%%%%%%%%%%%%%%%%%%%%%%%%%%%%%%%%%%%%%%%%%%%%%%%%%%%%%%%%%%%%%%%%%%%%
peaks in Fig.~\ref{fig:fom}. Specifically, near the Fermi energy $E_F\sim \pm 0.8$, $ZT\sim 8$, which is much larger than unity and certainly is a favorable TE response. Thus, we can say that such high values of $ZT$ are achieved by the enhancement of thermopower and the suppression of thermal conductance due to electrons in presence of strain, particularly for the stretched condition.

The effectiveness of the TE response is usually measured by the dimensionless quantity $ZT$ as discussed earlier. However, there are other key parameters to gauge the TE response of a material. Such as the power factor and efficiency, the studies which are also important to see the applicability of a TE material, especially for power generation applications~\cite{liu2016importance}. Therefore, below we discuss the power factor and the efficiency of pentacene both for the strain-less and strained cases.

\begin{figure}[ht!]
\centering
\includegraphics[width=0.238\textwidth]{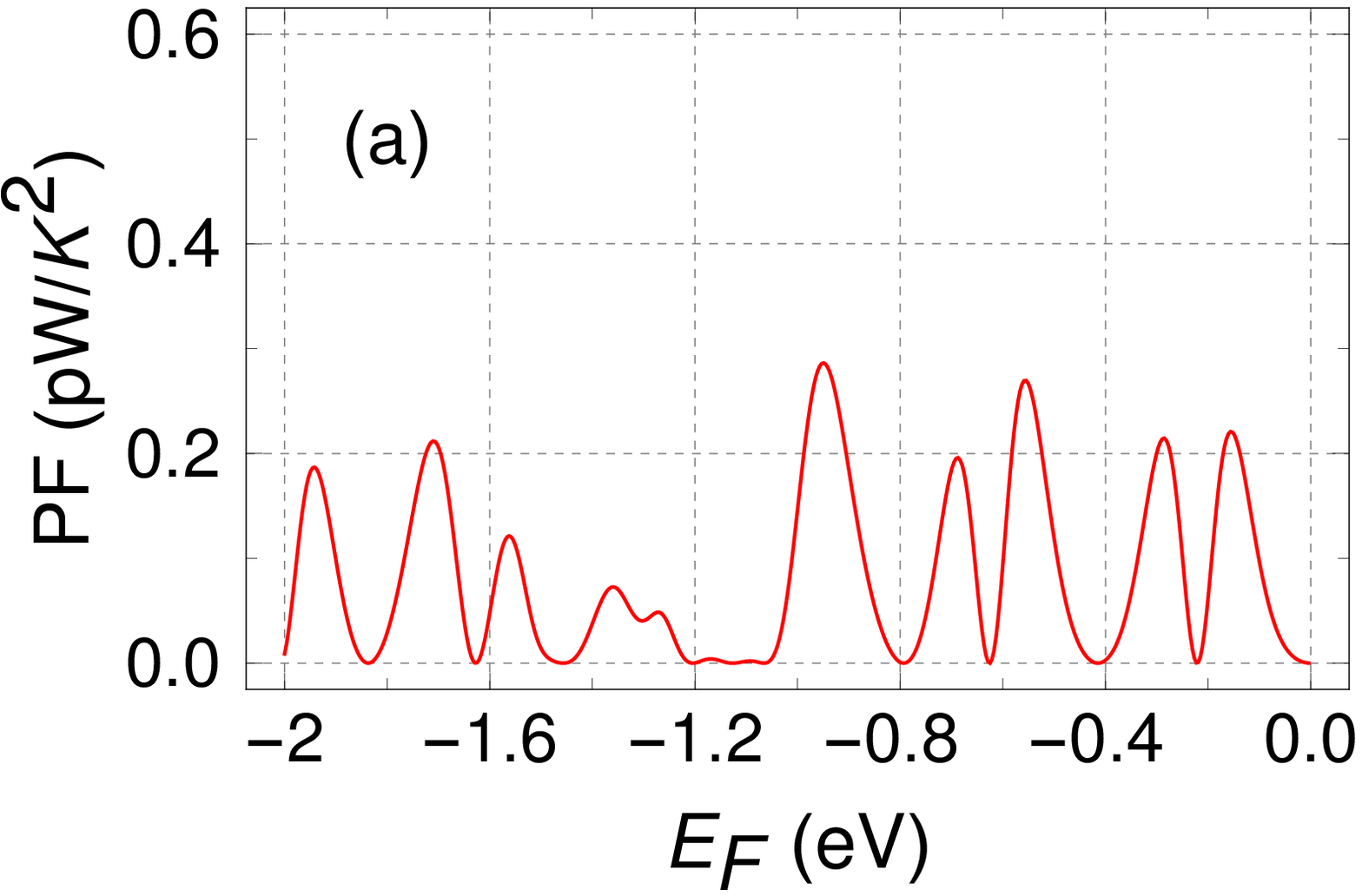} \vskip 0.1in
\includegraphics[width=0.238\textwidth]{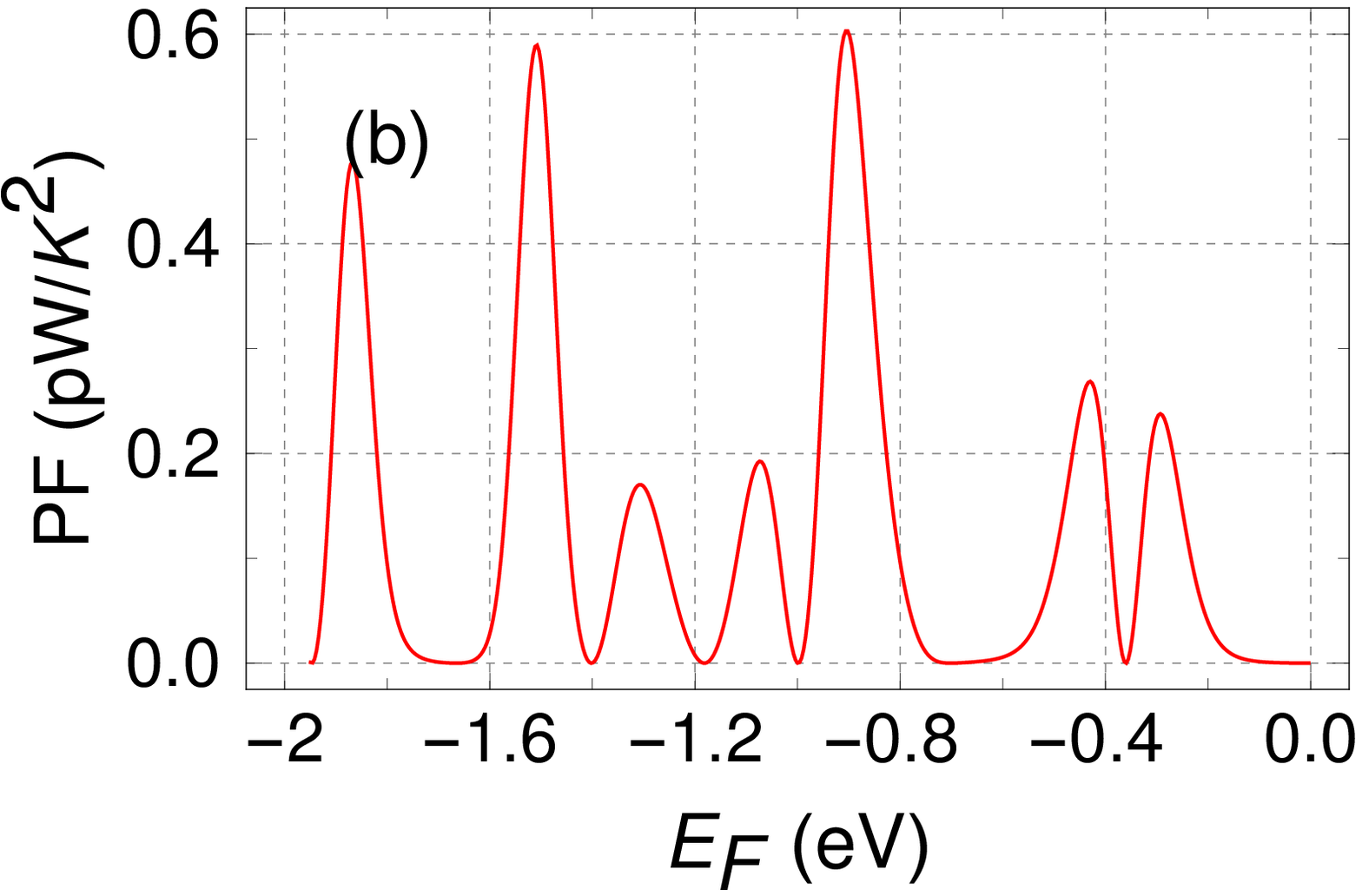} \hfill
\includegraphics[width=0.238\textwidth]{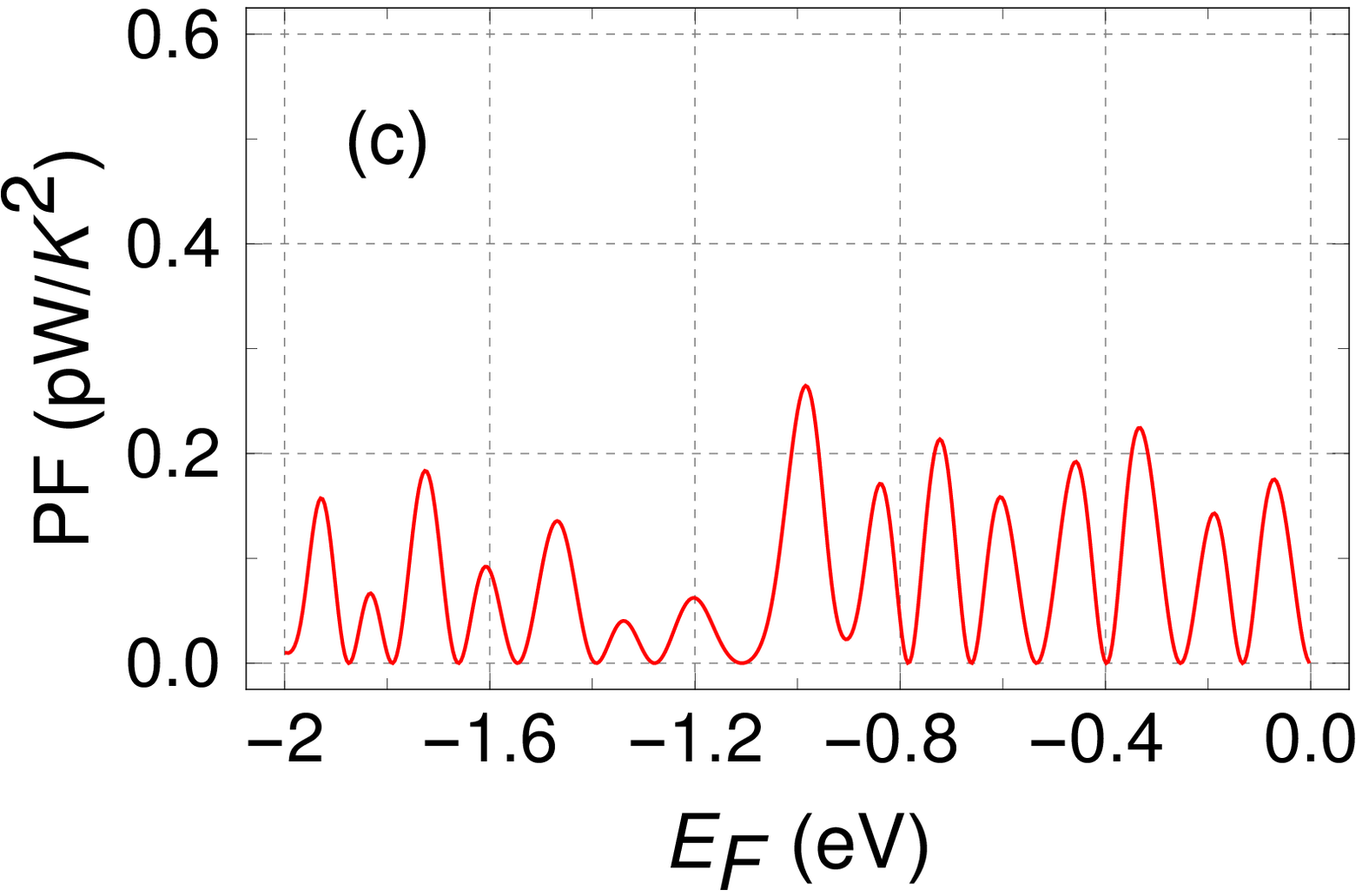} 
\caption{(Color online). Power factor $PF$ as a function of Fermi energy at room temperature for the (a) pristine case ($\delta =0$), (b) $\delta =0.3$, and (c) $\delta =-0.3$.}
\label{fig:pf}
\end{figure}

The behavior of the power factor as a function of Fermi energy is depicted in Fig.~\ref{fig:pf} for the $\delta$-values as considered before. For the strain-less case, the power has a maximum value $~0.3\,$pW/K$^2$. The behavior of the power factor for $\delta =- 0.3$ is more or less similar to the pristine case. 
%################################################################
%%%%%%%%%%%%%%%%%%%%%%%%%%%%%%%%%%%%%%%%%%%%%%%%%%%%%%%%%%%%%%%
However, the power factor for $\delta=0.3$ is almost twice the previous two cases and has a value of $~0.5$ pW/k$^2$ near $E_F\sim 0.8$. Thus, we see that the introduction of strain improves the power factor. However, the maximum power can be estimated by employing Eq.~\ref{eqn:maxP}. For example, at room temperature for $\delta =0.3$ and $E_F = -1.6$ eV, the maximum power is $0.125 (\Delta T)^2\,$pW/K$^2$, where $\Delta T$ is infinitesimal temperature difference between the electrodes.

\begin{figure}[ht!]
\centering
\includegraphics[width=0.238\textwidth]{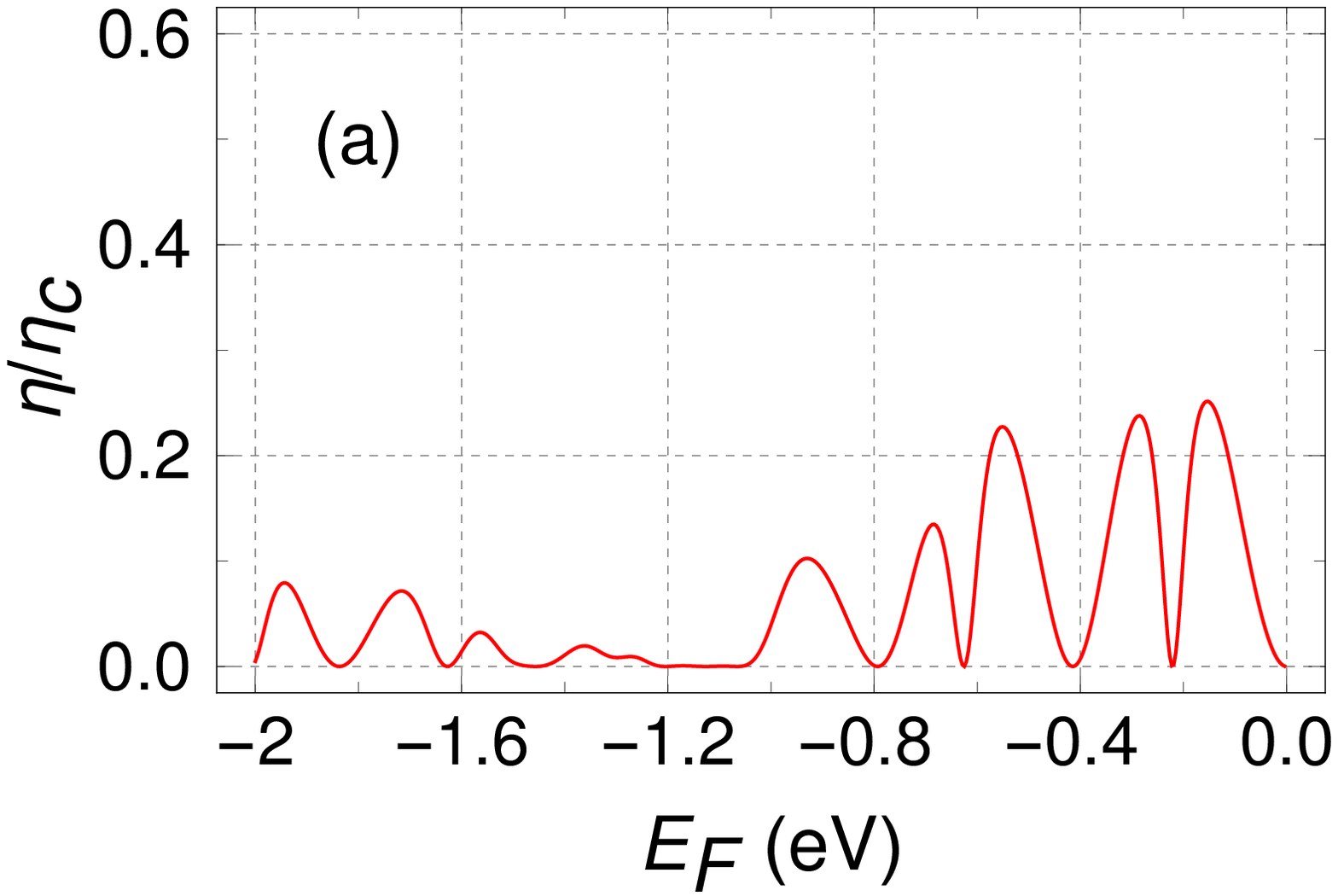} \vskip 0.1 in
\includegraphics[width=0.238\textwidth]{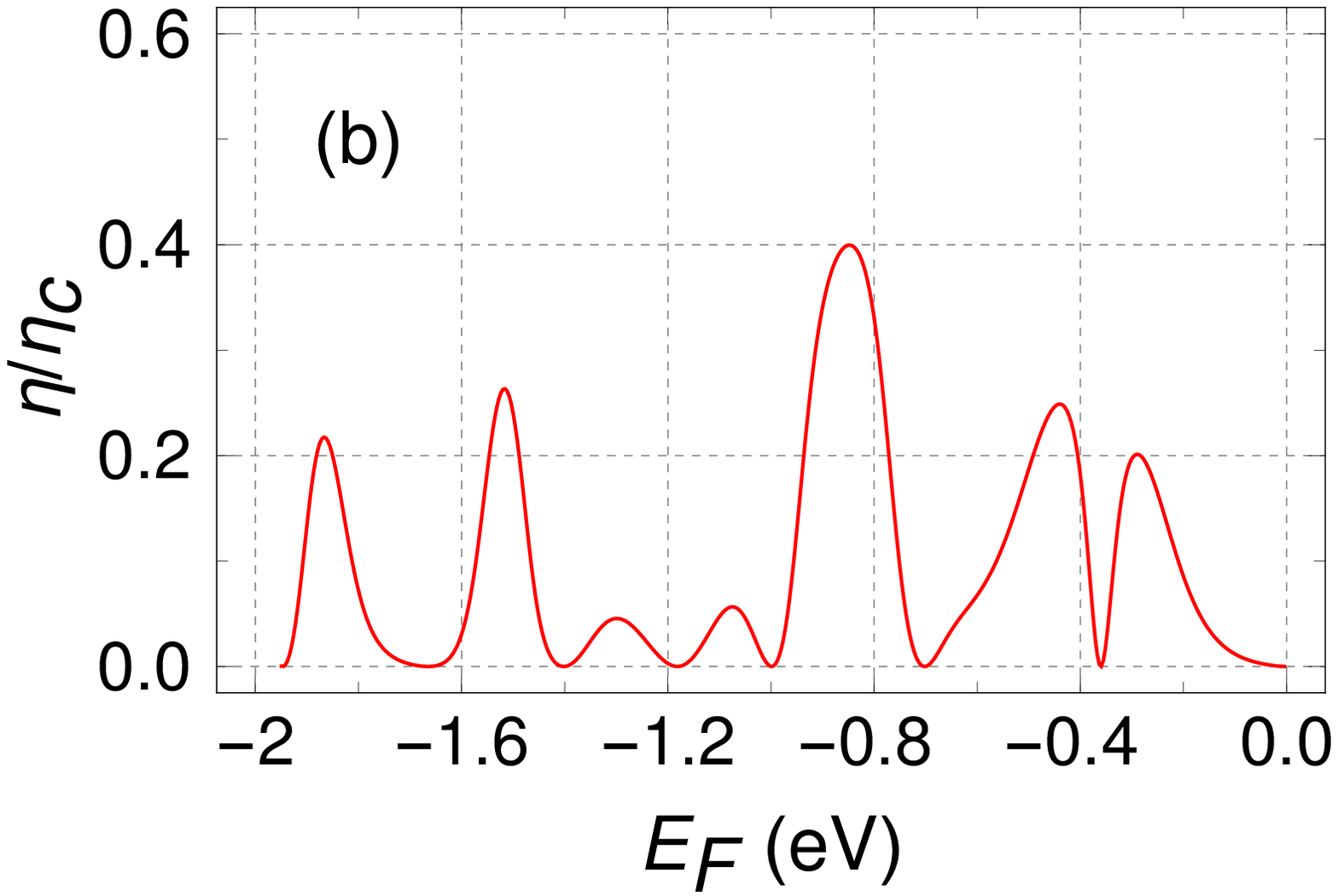} \hfill
\includegraphics[width=0.238\textwidth]{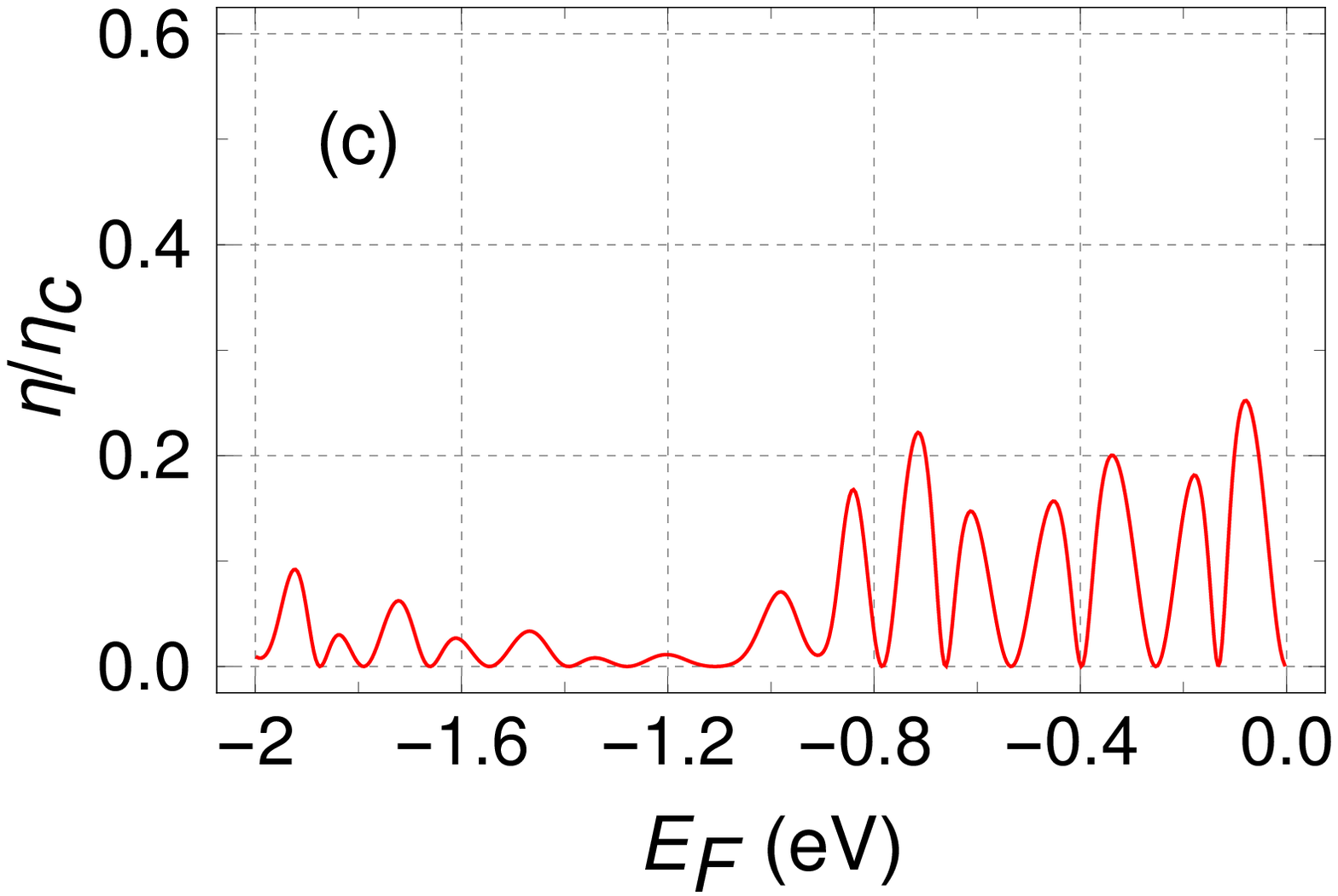} 
\caption{(Color online). Efficiency at maximum power $\eta/\eta_C$ as a function of Fermi energy at room temperature for the (a) pristine case ($\delta =0$), (b) $\delta =0.3$, and (c) $\delta =-0.3$.}
\label{fig:effi}
\end{figure}

It is easy to see from Eq.~\ref{eqn:etamaxP} that the efficiency at maximum power is just a parametric extension of \textit{figure of merit}. Naturally, we expect similar behavior of efficiency at maximum power to the $ZT$. The efficiency at maximum power ($\eta\left(P_{max}\right)$) (normalised to ideal Carnot efficiency $\eta_C$) is presented as a function of Fermi energy in Fig.~\ref{fig:effi}. 
The efficiencies at maximum power for $\delta=0$ (Fig.~\ref{fig:effi}(a)) and -0.3 (Fig.~\ref{fig:effi}(c)) are somewhat similar and the maximum efficiency is noted about 20$\%$ of the ideal Carnot engine efficiency. On the other hand, 
%################################################################
%%%%%%%%%%%%%%%%%%%%%%%%%%%%%%%%%%%%%%%%%%%%%%%%%%%%%%%%%%%%%%%
the efficiency at maximum power for $\delta = 0.3$ at room temperature is found to be $~0.4 ~\eta_C$ (Fig.~\ref{fig:effi}(b)), which is quite good from the application point of view.

\subsection{Thermal conductance due to phonon} 
Now, we discuss the results due to phonons. To begin with, let us mention the values that are considered in the present work. The system is treated as the spring-mass system to compute $k_{ph}$. As mentioned earlier, the spring constants are calculated from the second derivative of the harmonic Harrison potential~\cite{harrison}. For the 1D electrodes, we consider two different materials, namely, Ge and Au. The spring constant for Ge is considered as~\cite{aghosh} $13.71\,$N/m, while for Au~\cite{PhysRev.111.707}, it is $14.68\,$N/m. For the pentacene, the spring constant is $5.1\,$N/m in the absence of strain, which is considered as same as the single-crystal benzene~\cite{doi:10.1063/1.1725566}. The cut-off frequency of vibration depends on the material properties of the electrodes and the pentacene since we assume that two different atoms are adjacent to each other at the interface. For the chosen setup, by averaging the spring constants of the electrodes and pentacene, and the masses, the cut-off frequency for Ge electrodes comes out to be $\omega_c= 23.2\,$Trad/s, and for Au electrode, it is 15.1$\,$Trad/s. It should be noted that here the effect of strain is also considered through the change in bond length ($\delta=0.3$ case, which yielded the favorable TE response) as discussed earlier and the effective spring constant for pentacene is included in the calculation of the phonon thermal conductance.
\begin{figure}[h!]
\centering
\includegraphics[width=0.238\textwidth,height=0.15\textwidth]{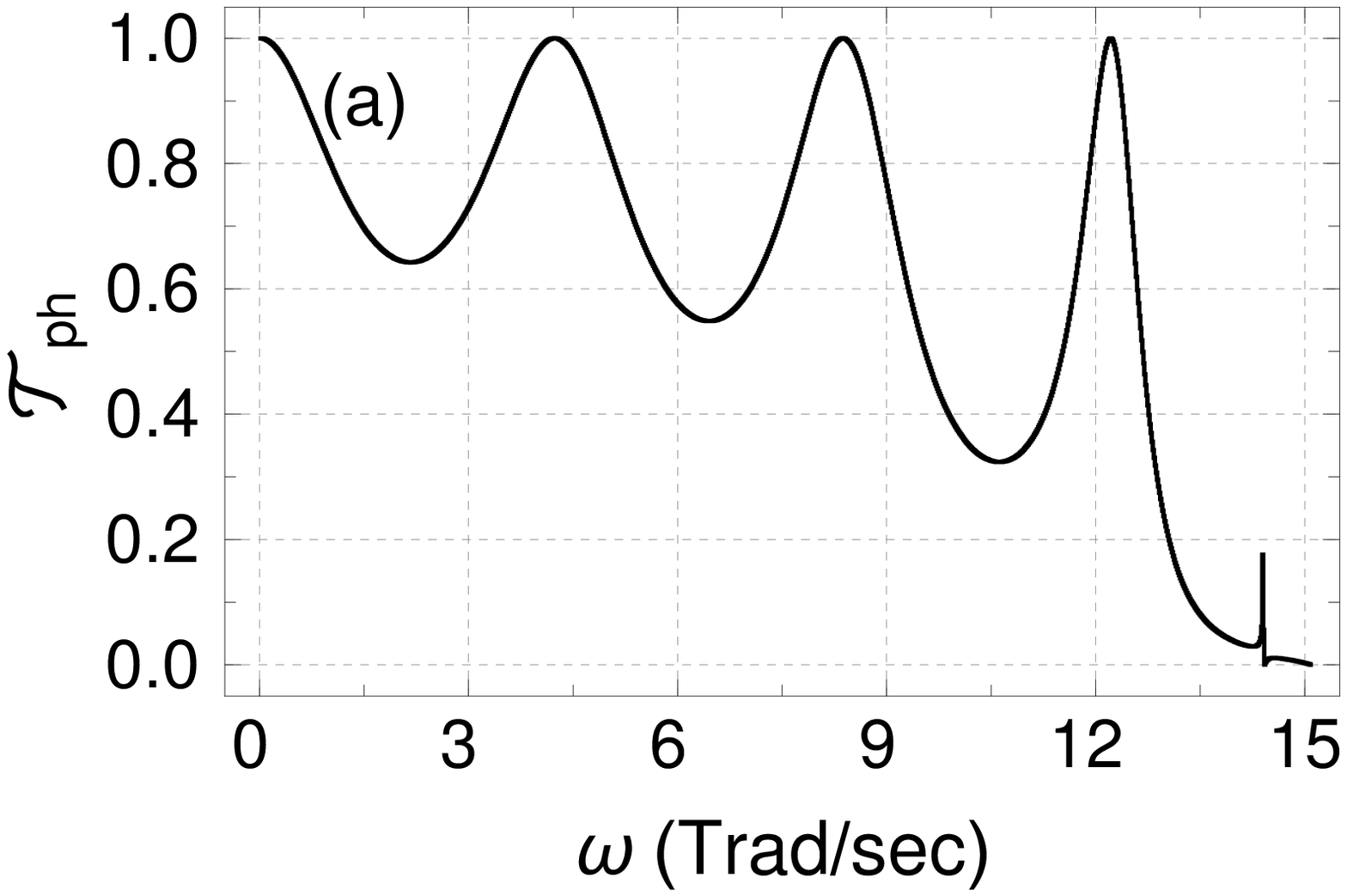} \hfill
\includegraphics[width=0.238\textwidth,height=0.15\textwidth]{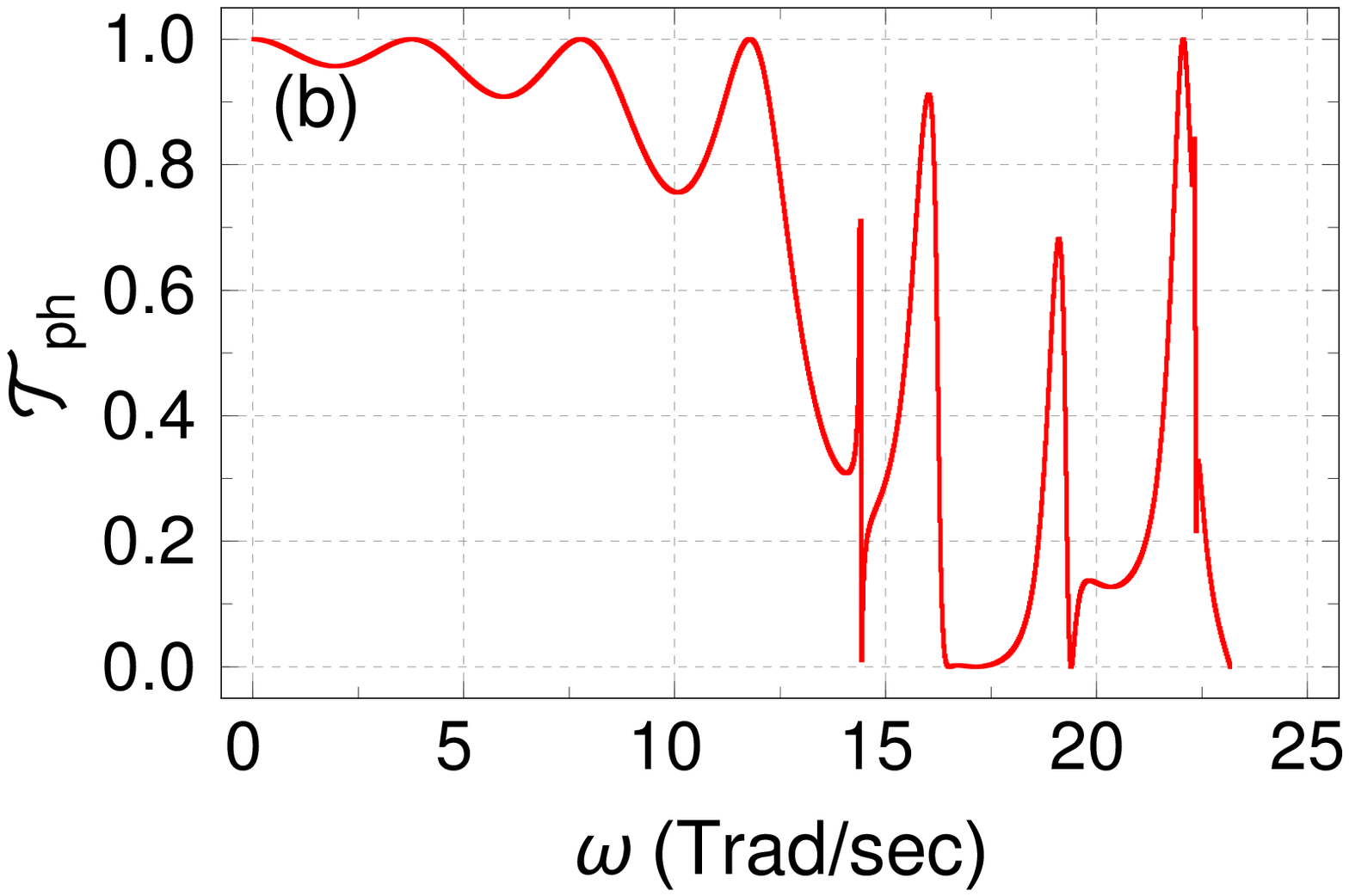} \vskip 0.1in
\includegraphics[width=0.238\textwidth,height=0.15\textwidth]{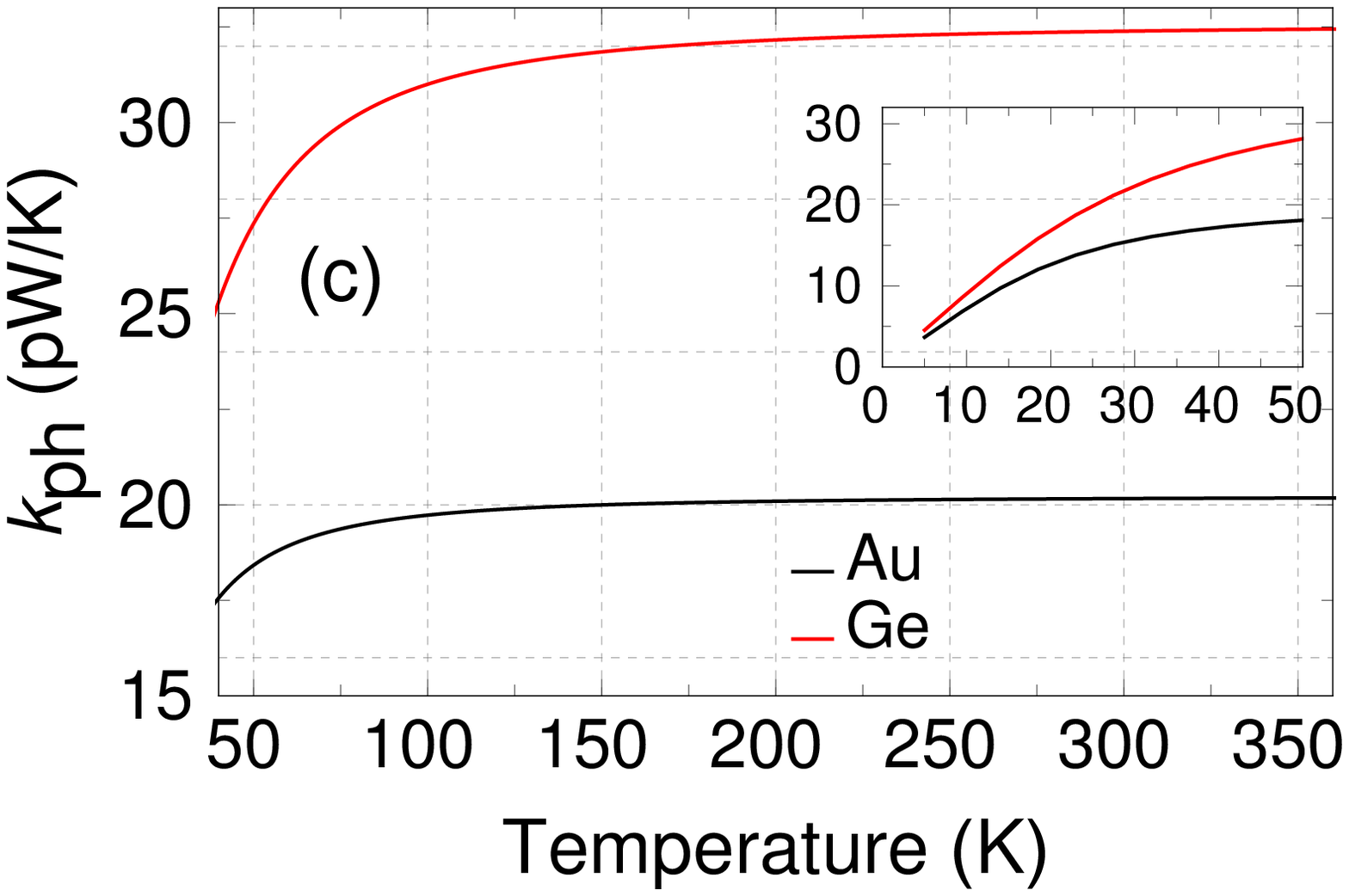} \hfill
\includegraphics[width=0.238\textwidth,height=0.15\textwidth]{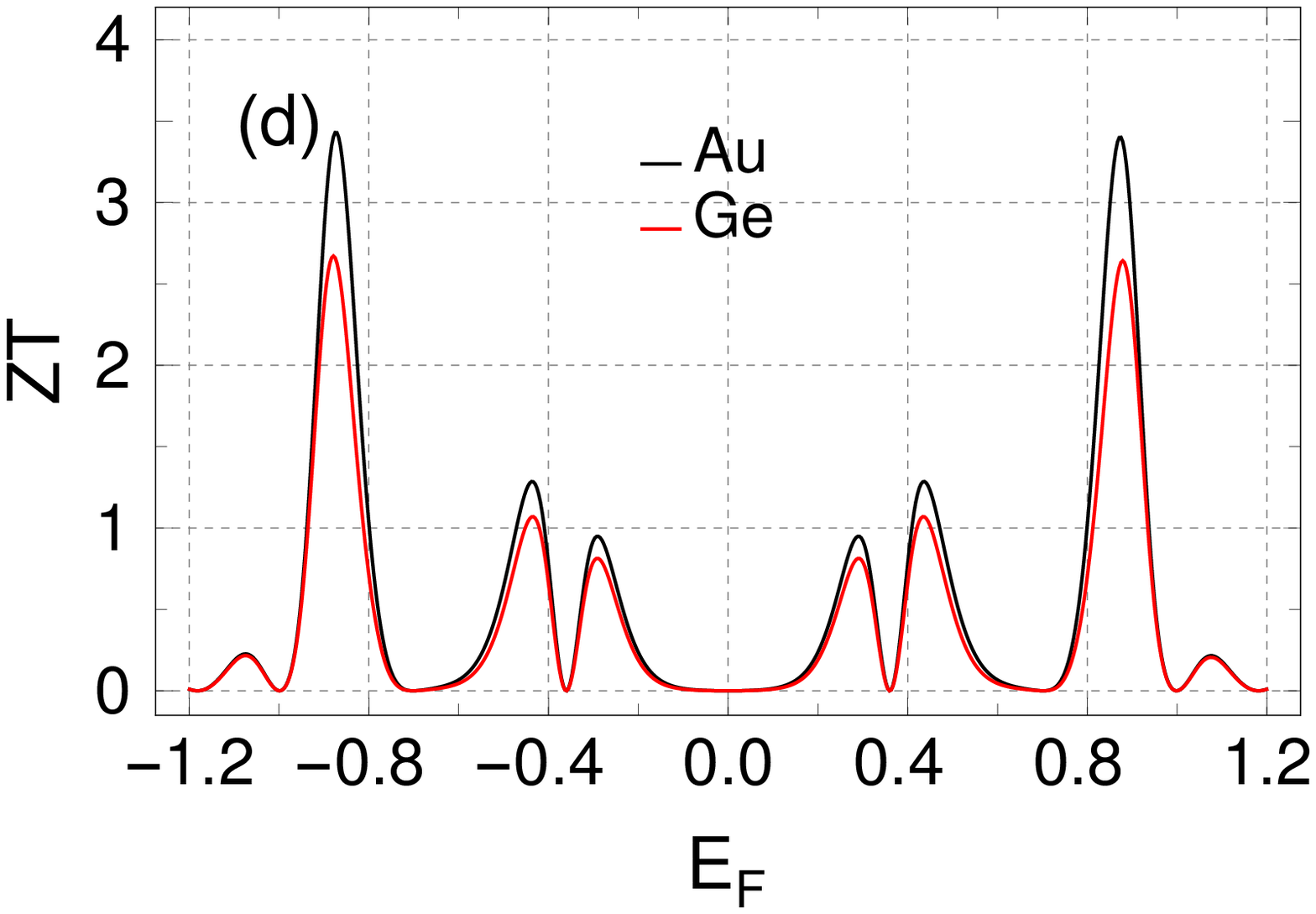} 
\caption{(Color online). Phonon transmission probability ${\mathcal T}_{ph}$ as a function of frequency $\omega$ for (a)  Gold(Au), and (b) Germenium(Ge). (c) Phonon thermal conductance $\kappa_{ph}$ as a function of temperature $T$ and the inset shows the low temperature behavior for those to electrodes. (d) $ZT$ as a function of Fermi energy at room temperature. All the results shown abve are in presence of strain with $\delta=0.3$.}
\label{fig:kph}
\end{figure}
In Figs.~\ref{fig:kph}(a) and (b), we plot the phonon transmission probability as a function of phonon frequency for the Au and Ge electrodes. Fabry-p\'{e}rot-likes peaks are observed in both the cases as discussed in Ref.~\cite{aghosh}. The behavior of phonon thermal conductance with temperature is shown in Fig.~\ref{fig:kph}(c). In the inset of the same plot, we show $k_{ph}$ variation at low temperatures. We observe that $k_{ph}$ is only a few pW/K at very low temperature, which is in good agreement with our previous statement~\cite{PhysRevB.102.245412}. As the temperature increases, $k_{ph}$ tends to saturate. The saturated value for Ge electrode is about 20$\,$pW/K and for Au electrode, it is about 32$\,$pW/K. With these values of $k_{ph}$, we finally compute the total $ZT$ which includes the thermal conductance due to electronic contribution as well as the phonons, as shown in Fig.~\ref{fig:kph}. The maximum $ZT$ is noted for Au electrode is about 3.5 and for Ge about 2.8, which are greater than unity. Thus it is evident that even if one includes the phononic contribution, favorable TE response persists at room temperature.

\subsection{Effect of Temperature}
Since, our study reveals that $\delta = 0.3$, that is, stretching the functional element is more effective for achieving the favorable TE response, 
%################################################################
\begin{figure}[ht!]
\centering
\includegraphics[width=0.495\textwidth]{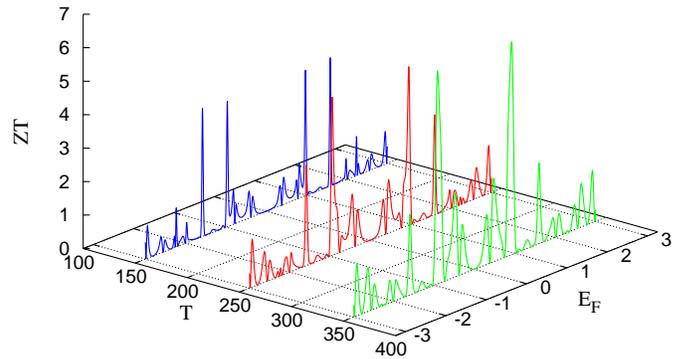} 
\caption{(Color online). $ZT$ as a function of $E_F$ at three different temperatures, namely, $T=150\,$K (blue), 250$\,$K (red), and 350$\,$K (green) for $\delta =0.3$.}
\label{fig:temp}
\end{figure}
%%%%%%%%%%%%%%%%%%%%%%%%%%%%%%%%%%%%%%%%%%%%%%%%%%%%%%%%%%%%%%%%%%%%%%
we examine the temperature dependence of $ZT$ for this particular case. In Fig.~\ref{fig:temp}, we plot $ZT$ for three different temperature values, namely, $150\,$K, $250\,$K, and $350\,$K, represented by blue, red, and green colors, respectively. Here, $ZT$ tends to increase with temperature and the maximum values of $ZT$ for the said temperatures within the given Fermi energy window are about $3$, $6$, and $9$, respectively. Note that within a wide range of temperatures, $ZT$ is always higher than unity.

\subsection{Effect of electrode-molecule coupling}
The electrodes-molecule coupling strength plays a crucial role in TE response. In all the earlier plots, the coupling strength was fixed at 0.75$\,$eV. Here we also consider two other different strengths, namely, $\tau=0.5\,$eV and 1$\,$eV to study the TE performance. It should be noted that all the chosen coupling strengths are within the wideband limit. The results for $\tau=0.5\,$eV, 0.75$\,$eV, and 1$\,$eV are denoted with blue, green, and red colors, respectively. Here $\tau=0.75\,$eV is included for the sake of comparison. First we study the transmission 
\begin{figure}[ht!]
\centering
\includegraphics[width=0.238\textwidth,height=0.16\textwidth]{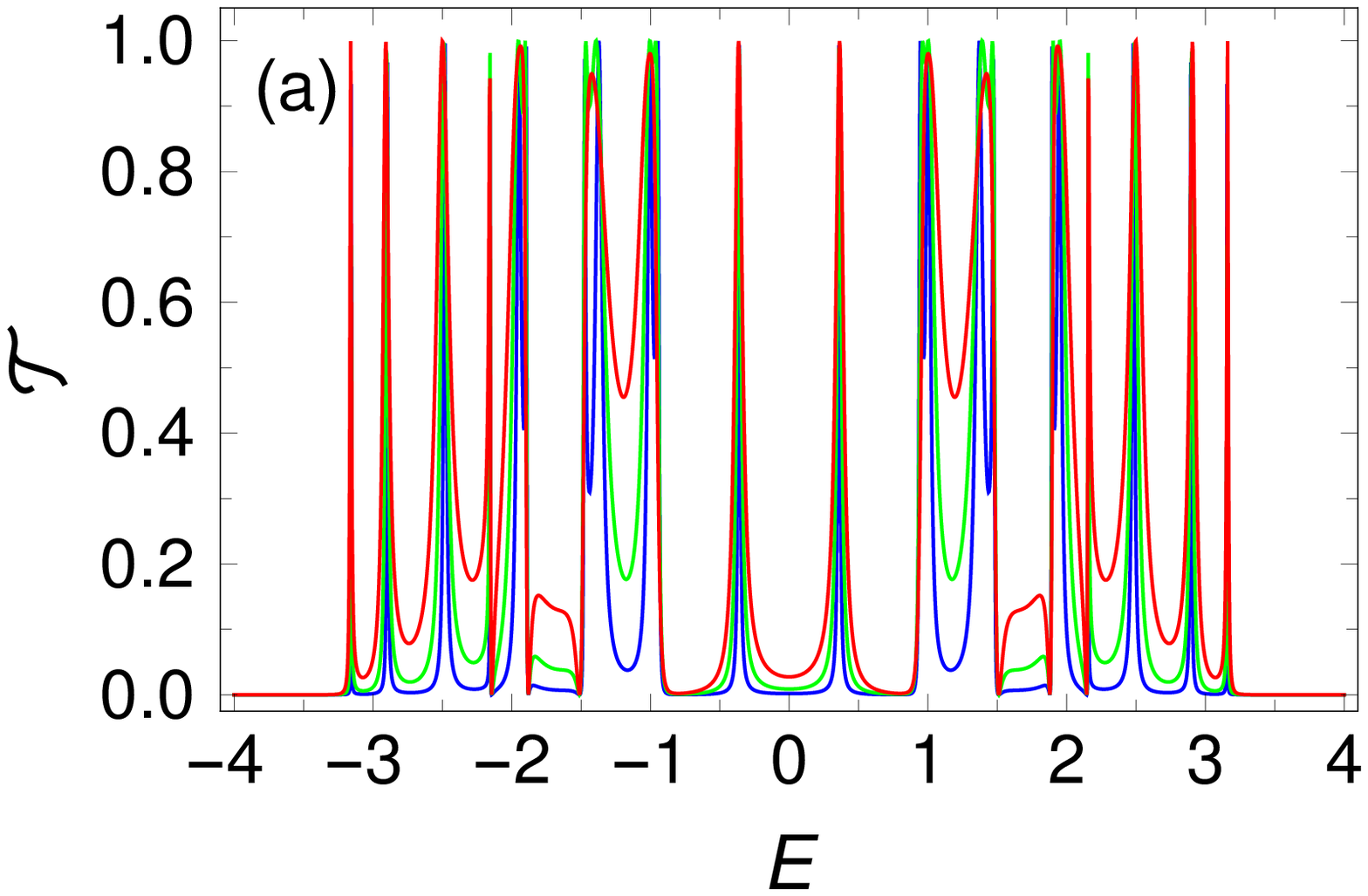} \hfill \includegraphics[width=0.238\textwidth,height=0.16\textwidth]{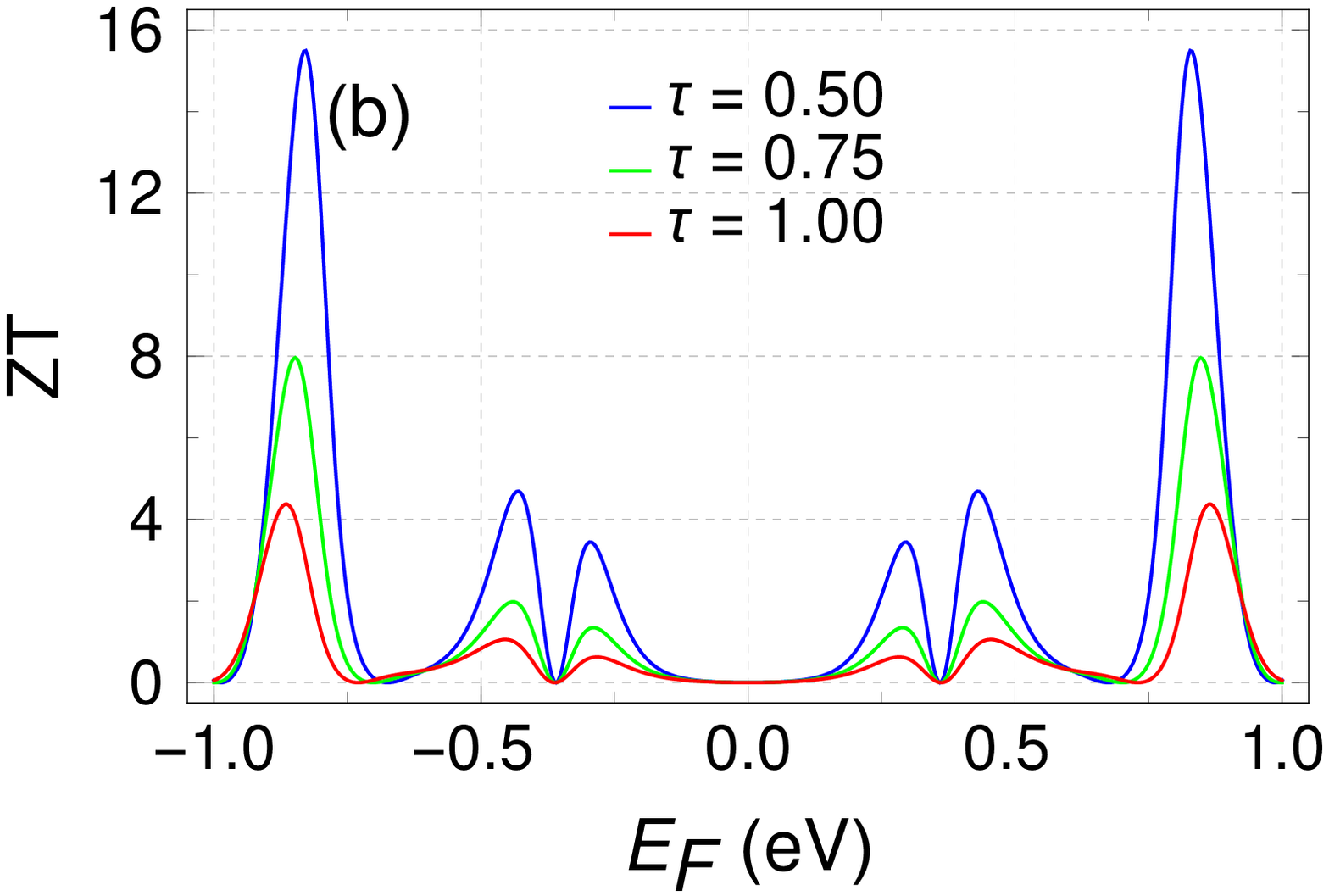} 
\caption{(Color online). (a) Transmission probability as a function of energy, (b) $ZT$ as a function of $E_F$ for three different coupling strengths, namely $\tau=0.5\,$eV (blue), 0.75$\,$eV (green), and 1$\,$eV (red). The results corresponding to $\tau=0.75\,$eV is included for the sake of comparison.}
\label{fig:coup}
\end{figure}
spectrum as shown in Fig.~\ref{fig:coup}(a). The coupling strengths between the electrodes and molecule provide the broadening in the transmission spectrum. For instance, in the case of weak coupling strength ($\tau=0.5\,$eV, blue curve), the transmission probability shows a few sharp peaks. Other than the sharp peaks, transmission probability is either zero or vanishingly small throughout the given energy window. On the other hand, for the strong coupling strength ($\tau=1\,$eV, red curve), the transmission spectra is comparatively broadened than the other two cases. The transmission function becomes more asymmetric with reducing the electrode-molecule coupling strength. Naturally, much favorable response is expected in the limit of weak-coupling, which is clearly reflected from Fig.~\ref{fig:coup}. From weak to strong coupling strengths, the value of $ZT$ decreases in a regular manner. For $\tau=0.5\,$eV, the maximum $ZT$ is noted about 16, while $ZT\sim 4$ for $\tau=1\,$eV. Nevertheless, all the chosen coupling strengths exhibit favorable TE responses.

\subsection{Role of substrate}
Usually, molecules are grown on glass or metallic substrates~\cite{PhysRevB.69.075408,Fritz2004,Shioya2019}. Recently, pentacene has also been prepared by vacuum deposition with ionic liquids~\cite{Takeyama2011}. Since the substrates or the immediate environment around the molecule can modify the electronic properties of the molecule, their effects must be taken into account while studying the TE response. Such effects can be incorporated via disorder in the present case, which may arise due to the trapped charge impurities on the surface, surface roughness, etc.~\cite{Rycerz_2007,Lewenkopf2013}. Therefore, we consider disorder 
\begin{figure}[ht!]
\centering
\includegraphics[width=0.3\textwidth]{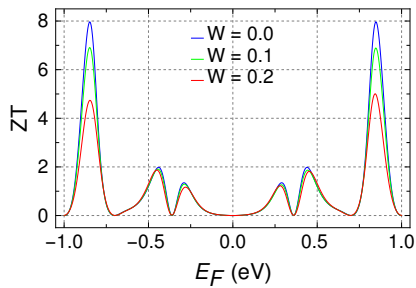} 
\caption{(Color online). $ZT$ as a function of $E_F$ at three disorder strengths, namely, $W=0$ (blue), 0.1 (green), and 0.2 (red) for $\delta =0.3$.}
\label{fig:dis}
\end{figure}
effect in our system, where the disorders are included in the site energies of the pentacene `randomly' from a Box distribution function of width $W$, that is the site energies are chosen randomly within the range $[-W/2:W/2]$. As the site energies are random, we compute
$ZT$ by taking the average of over a large number of distinct disordered configurations, to get reliable results. In Fig.~\ref{fig:dis}, we plot $ZT$ as a function of $E_F$ for two disorder strengths, namely, $W=0.1$ (denoted with green) and 0.2 (denoted with red), and include the disorder-free case (denoted with blue) for comparison. In presence of disorder, the transmission probability will have lower values due to the localization effect and hence the degree of TE performance will suppress. This fact is clearly seen in Fig.~\ref{fig:dis}, specifically, near the Fermi energy $E_F\sim \pm1\,$eV. The important observation is that for $W=0.2$, $ZT$ is still greater than unity, which is a favorable response.

\subsection{Effect of dangling bonds}  
Finally, we study the effect of dangling bonds on the TE performance, which may present in pentacene. Usually, hydrogen (H) bonding or H-replacement is not energetically favored in pentacene on H-passivated surfaces\cite{doi:10.1063/1.2139989}. However, when a single pentacene molecule is chemically absorbed on silicon surface, the molecule may acquire a few dangling bonds with H as well as Si~\cite{CHOUDHARY200520,SUZUKI20065092}. Therefore, in order to see whether the presence of dangling bonds enhances the TE performance, we consider only the dangling bonds with H atoms~\cite{doi:10.1143/JPSJ.68.1321}. Here we also consider the strain condition $\delta=0.3$, which gave us the favorable TE response. The hopping integral between the sites of the pentacene and H is considered as 0.9$\,$eV. The on-site energy for the dangling sites is set at 0.1$\,$eV. 
\begin{figure}[ht!]
\centering
\includegraphics[width=0.3\textwidth]{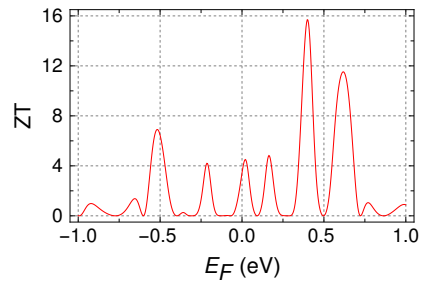} 
\caption{(Color online). $ZT$ as a function of $E_F$ in presence of dangling bonds for $\delta =0.3$.}
\label{fig:dang}
\end{figure}
All other parameters are identical as earlier. In Fig.~\ref{fig:dang}, we plot $ZT$ as a function of the Fermi energy $E_F$ considering the dangling bonds. Interestingly, the result is highly favorable, as the maximum $ZT$ is noted at about 16. This considerable enhancement in $ZT$ is quite expected. This is because, as we introduce the dangling bonds in our system, we basically include some asymmetry in the system through the on-site energies and the hopping integrals in the dangling bonds. This, in turn, makes the transmission function more asymmetric and hence the significant enhancement in $ZT$.

Here it is relevant to note that other $\delta$-values can also be taken into account to investigate the TE properties. From our exhaustive calculations, we find favorable responses for other values of $\delta$ as well.

\vskip 0.2 cm

{\it Experimental feasibility}: Recently, the solution processing techniques have been used to alter the molecular packing through lattice strain leading to efficient, low-cost organic semiconductor devices~\cite{Zhenan,Sakanoue2010}. For example, using solution shearing methods, Giri {\it et al}. have shown that $\pi -\pi$ molecular stacking is reduced from $3.33^\circ$ to $ 3.03^\circ$ in the case of TIPS pentacene which is a derivative of pentacene~\cite{Giri2011}. One of the advantages of the above-mentioned solution-shearing technique is that it is possible to access new packing structures that are not possible through other commonly used methods, which increases the possibility of exploring the effects of the mechanical strain in a more general way.

\section{\label{sec:conclusion}Concluding remarks }
%%%%%%%%%%%%%%%%%%%%%%%%%%%%%%%%%%%%%%%%%%%%%%%%%%%%%%%%%%%%%%%%%%%%%%
In the present work, we have proposed a scheme to achieve a favorable TE response in a typical small organic molecule, namely, pentacene, via a non-synthetic strategy. We have particularly shown that the TE performance can be enhanced significantly when the system under consideration is subjected to a uni-axial strain along the longitudinal direction. The proposed setup is described within the nearest-neighbor tight-binding framework and the effect of uni-axial strain is incorporated through the hopping integrals. Different thermoelectric quantities are computed using Green's function formalism following the Landauer-B\"{u}ttiker prescription.

The key outcomes of the present work are as follows:

$\bullet$ The uni-axial strain induces a spatial anisotropy in the hopping integrals, which significantly modifies the transmission spectra.

$\bullet$ The thermopower increases considerably and the thermal conductance due to electrons is suppressed significantly for the stretched condition than those for the pristine and compressed cases.

$\bullet$ Corresponding $ZT$ is much higher than unity at room temperature and thus is a favorable response.

$\bullet$ The studies of power factor and efficiency at maximum power show that our proposed setup is suitable for application purposes.

$\bullet$ The inclusion of the phononic contribution though suppresses the TE response, $ZT$ is still higher than unity.

$\bullet$ $ZT$ shows an increasing trend with temperature.

$\bullet$ The role of substrate, presence of dangling bonds, modulation of the electrode-molecule coupling strength, each of the scenarios exhibits favorable TE response.
 
In view of the promising results of the present work, our proposal provides a significant boost to the TE community. Various simple/complex molecular structures can be explored via our proposed non-synthetic strategy, that is, strain. The said exploration may yield non-trivial TE responses and may find suitable for designing efficient TE devices.

%%%%%%%%%%%%%%%%%%%%%%%%%%%%%%%%%%%%%%%%%%%%%%%%%%%%%%%%%%%%%%%

\section*{COMPETING INTERESTS STATEMENT}

The authors declare that they have no known competing financial interests or personal relationships that could have appeared to influence the work reported in this paper.

%%%%%%%%%%%%%%%%%%%%%%%%%%%%%%%%%%%%%%%%%%%%%%%%%%%%%%%%%%%%%%%

\section*{DATA  AVAILABILITY STATEMENT }

Derived data supporting the findings of this study are available from the authors on request.

%%%%%%%%%%%%%%%%%%%%%%%%%%%%%%%%%%%%%%%%%%%%%%%%%%%%%%%%%%%%%%%%%

\section*{AUTHOR CONTRIBUTION STATEMENT}

KM, SG, and SKM conceived the project. KM and SG performed the numerical calculations. KM, SG, and SKM  analyzed the data and co-wrote the paper.

%%%%%%%%%%%%%%%%%%%%%%%%%%%%%%%%%%%%%%%%%%%%%%%%%%%%%%%%%%%%%
%         							References
%%%%%%%%%%%%%%%%%%%%%%%%%%%%%%%%%%%%%%%%%%%%%%%%%%%%%%%%%%%%%

\bibliography{ref}
\bibliographystyle{apsrev4-1}

%%%%%%%%%%%%%%%%%%%%%%%%%%%%%%%%%%%%%%%%%%%%%%%%%%%%%%%%%%%%%
\end{document}